\documentclass[journal,draftcls,onecolumn,12pt,twoside]{IEEEtranTCOM}
\normalsize
\usepackage{cite}
\usepackage{amsmath,amssymb,amsfonts,mathtools}
\usepackage{algorithmic}
\usepackage{graphicx}
\usepackage{textcomp}
\usepackage{xcolor}
\usepackage{psfrag}
\usepackage{float}
\usepackage{subcaption}
\usepackage{amsmath}

\usepackage[skip=0pt]{caption}
\usepackage{wrapfig}
\usepackage[utf8]{inputenc}

\usepackage{multirow}
\usepackage{subcaption}

\newcommand{\norm}[1]{\left\|#1\right\|}

\newcommand{\f}[1]{\boldsymbol{#1}}

\newcommand{\s}[1]{\mathsf{#1}}

\DeclareMathOperator*{\argminA}{arg\,min}
\onecolumn

\begin{document}
\title{Low PAPR MIMO-OFDM Design Based on Convolutional Autoencoder}

\author{Yara~Huleihel, and~Haim~H.~Permuter,~\IEEEmembership{Senior~Member,~IEEE}\vspace{-1cm}}

\maketitle

\begin{abstract}
An enhanced framework for peak-to-average power ratio ($\mathsf{PAPR}$) reduction and waveform design for Multiple-Input-Multiple-Output ($\mathsf{MIMO}$) orthogonal frequency-division multiplexing ($\mathsf{OFDM}$) systems, based on a convolutional-autoencoder ($\mathsf{CAE}$) architecture, is presented. The end-to-end learning-based autoencoder ($\mathsf{AE}$) for communication networks represents the network by an encoder and decoder, where in between, the learned latent representation goes through a physical communication channel. 
We introduce a joint learning scheme based on projected gradient descent iteration to optimize the spectral mask behavior and MIMO detection under the influence of a non-linear high power amplifier ($\mathsf{HPA}$) and a multipath fading channel. The offered efficient implementation novel waveform design technique utilizes only a single $\mathsf{PAPR}$ reduction block for all antennas. It is throughput-lossless, as no side information is required at the decoder. 
Performance is analyzed by examining the bit error rate ($\mathsf{BER}$), the $\mathsf{PAPR}$, and the spectral response and compared with classical $\mathsf{PAPR}$ reduction $\mathsf{MIMO}$ detector methods on 5G simulated data. The suggested system exhibits competitive performance when considering all optimization criteria simultaneously.
We apply gradual loss learning for multi-objective optimization and show empirically that a single trained model covers the tasks of $\mathsf{PAPR}$ reduction, spectrum design, and $\mathsf{MIMO}$ detection together over a wide range of SNR levels.
\end{abstract}

\vspace{-0.6cm}
\section{Introduction}
The Multiple-input-multiple-output ($\mathsf{MIMO}$) scheme is a widely used technique for enhancing channel capacity and transmission reliability, thanks to the diversity and multiplexing gains. Orthogonal frequency division multiplexing ($\mathsf{OFDM}$) is a waveform design method known for providing high bandwidth efficiency, high throughput, simple equalization in wireless transmission, and efficient hardware implementation. For these reasons, it has been adopted as a standard technology in various wireless communication systems, such as WiFi, 4G, and 5G standards for wireless communications.
Nonetheless, significant drawbacks of the $\mathsf{OFDM}$ multi-carrier system appear in the form of adjacent channel power ratio ($\mathsf{ACPR}$) limitations and the tendency to produce signals with a high peak-to-average power ratio ($\mathsf{PAPR}$) in the time-domain, since many subcarrier components are added via a fast Fourier transform ($\mathsf{FFT}$) operation. The contribution of each subcarrier to the total power is dynamic, which makes the total power highly variable. In particular, the high $\mathsf{PAPR}$ in $\mathsf{MIMO}$-$\mathsf{OFDM}$ systems is exacerbated as the number of antennas increases \cite{block_papr}. The demand for higher energy efficiency as well as reduced power consumption is expected to increase for future radio systems \cite{6G_2}. Moreover, future communication system design and equipment are expected to be more compatible with machine learning ($\mathsf{ML}$) implementation requirements, for example, allowing learning in the field to make some design choices \cite{6G_2}. Accordingly, waveform designs using $\mathsf{ML}$ techniques are becoming significantly attractive.  

A high power amplifier ($\mathsf{HPA}$) is required to provide enough output power for reliable communication. In practical systems, the $\mathsf{HPA}$ is not linear and distorts the transmitted signal. As a result, severe non-linear signal distortions are found when these high $\mathsf{PAPR}$ signals pass through the non-linear $\mathsf{HPA}$. The resulting signal exhibits spectral regrowth in the form of in-band signal distortions and out-of-band radiation \cite{band_ref}, and the bit error rate ($\mathsf{BER}$) increases. Hence, it is crucial to develop $\mathsf{PAPR}$ reduction techniques for $\mathsf{MIMO}$-$\mathsf{OFDM}$ systems to increase their efficiency in handling large data streams and to reduce their error rates.
Training and applying the $\mathsf{PAPR}$ reduction block to each antenna individually exacerbates the complexity, cost, and redundancy in proportion to the number of transmit antennas in the system. Instead, in this work, a single $\mathsf{PAPR}$ reduction block jointly operates on all antenna $\mathsf{OFDM}$ sequences, and it is designed according to the maximum $\mathsf{PAPR}$ value of all antenna sequences.

A central difficulty of the multiple transmitter (TX) and receiver (RX) antenna structure is posed by the need for joint detection of the data symbols sent by each transmitter. Unfortunately, the optimal $\mathsf{MIMO}$ detection solution imposes an NP-hard problem on the receiver. Consequently, various sub-optimal yet feasible detection algorithms have been proposed. Other than classical model-driven solutions, an increasing effort has been dedicated to $\mathsf{ML}$, and, specifically, deep learning ($\mathsf{DL}$) based techniques to solve the $\mathsf{MIMO}$ detection problem, and more generally, various wireless communication tasks.      

The design of $\mathsf{OFDM}$ waveform signals aims to simultaneously achieve a high data rate, high spectral efficiency (measured by the $\mathsf{ACPR}$), and low computational complexity \cite{ai4b5g, cho}. This design is highly affected by the non-linear effects of the $\mathsf{HPA}$. While keeping the $\mathsf{PAPR}$ level low is favorable, it is of particular importance to have acceptable signal spectral behavior and $\mathsf{BER}$, which are often referred to as \emph{waveform design}. In order to fulfill that, this work suggests an overall communication network multi-objective optimization, such that the transmitter, $\mathsf{HPA}$, channel, and the receiver, are represented as a single optimization block. Instead of separately optimizing different components of the transmitter and the receiver, an end-to-end convolutional-autoencoder ($\mathsf{CAE}$) learning model is proposed. This end-to-end optimization block is presented as a constrained optimization problem where the transmitted signal estimation is the objective, and the $\mathsf{PAPR}$ and $\mathsf{ACPR}$ requirements are the constraints. $\mathsf{MIMO}$ detection over multiple channel realizations is performed as a part of the end-to-end joint optimization model, utilizing an iterative approach based on convolutional layers, and a gradual loss learning approach.      
We evaluate the performance of our algorithms over both additive white Gaussian noise ($\mathsf{AWGN}$) and 3rd Generation Partnership Project (3GPP) fading channels \cite{3gpp}. By analyzing the BER, PAPR, and spectrum performance, we show that the proposed end-to-end learning approach can integrate different communication network blocks to balance those performance objectives successfully. We show that the suggested scheme is able to achieve better spectral performance for higher $\mathsf{HPA}$ efficiency operation.  
Various $\mathsf{OFDM}$ $\mathsf{PAPR}$ reduction techniques have been proposed in the literature, as well as for $\mathsf{MIMO}$ detection. Generally, these techniques can be categorized into \emph{model-driven} and \emph{data-driven} techniques. The first category refers to standard approaches in classical communications theory, while the second relies on recently developed approaches based on $\mathsf{ML}$ techniques. The following subsections review different earlier solutions for the above-mentioned problems.
\vspace{-0.3cm}
\subsection{Classical Approaches (Model Driven) for MIMO Detection}
Many $\mathsf{MIMO}$ detection algorithms have been developed over the years. The maximum likelihood estimation ($\mathsf{MLE}$) solution is optimal for the joint detection of transmitted symbols in a $\mathsf{MIMO}$ system. However, its exponential computational and time complexity (due to exhaustive searches over all possible transmitted signals) render it infeasible when the number of transmitters and the modulation order are high. 
An example of suboptimal high accuracy non-linear detection algorithms are those based on sphere decoding (SD) \cite{sd_1}, but they become computationally expensive as the number of antennas grows. The general idea is based on a lattice search for a solution in an iterative manner, and the accuracy/complexity ratio strongly depends on the value chosen for the radius parameter. 
More advanced detectors include the successive interference cancellation (SIC) based detectors \cite{vblast} and the semi-definite relaxation detectors \cite{sdr}.
\vspace{-0.4cm}
\subsection{Machine-Learning-Based Schemes (Data Driven) for MIMO Detection}
The motivation for DL-based detectors is to enhance the performance of classical model-driven detection algorithms by learning, from the training data set, an optimized mapping of the received signals onto the transmitted symbols. In \cite{det_net, NN_detector}, an iteration-based algorithm for implementing a receiver for $\mathsf{MIMO}$ detection was suggested. One of the highlights of the presented model's framework is that it enables training through different random communication channel realizations. In \cite{sd_2}, a model-based algorithm was suggested, where a classical SD algorithm was integrated with a neural network ($\mathsf{NN}$) that was trained to optimize the selection of the initial radius. In \cite{RE_MIMO}, a neural detector-based transformer architecture implements a recurrent estimation scheme by learning an iterative decoding algorithm. 

In \cite{MIMO_det, AE_detect_1, AE_detect_2} an $\mathsf{AE}$ was offered to design a physical layer in which DL-based CSI encoding was suggested for different scenarios to achieve lower $\mathsf{BER}$ together with better robustness to the wireless channel characteristics.
Thorough surveys and analysis are presented in \cite{ MIMO_det_Surv_2}.
\vspace{-0.4cm}
\subsection{Classical Approaches (Model Driven) for PAPR Reduction}
\vspace{-0.3cm}
$\mathsf{PAPR}$ reduction schemes are roughly classified into three categories. The signal distortion category consists of techniques such as clipping and filtering ($\mathsf{CF}$) \cite{clip1,clipSLM}, which limit the peak envelope of the input signal in the time domain to a predetermined value. The multiple signaling probabilistic category includes methods such as selective mapping ($\mathsf{SLM}$) \cite{clipSLM, SLM_PTS}, partial transmit sequence ($\mathsf{PTS}$) \cite{SLM_PTS}, ton reservation and ton injection \cite{PAPR_OVER_2}, and constellation shaping \cite{cshape2}. The main principle of $\mathsf{SLM}$ is to generate different candidates for each $\mathsf{OFDM}$ block by multiplying the symbols vector with a set of different pseudo-random sequences and choosing the candidate with the lowest $\mathsf{PAPR}$. The third category is the coding technique category \cite{PAPR_OVER_2, PAPR_CODING}, attempting to reduce the occurrence probability of the same phase signals.

Earlier schemes were mainly developed for single-antenna systems. Extended works which applied the single-antenna $\mathsf{PAPR}$ reduction schemes on each antenna of the $\mathsf{MIMO}$ configuration separately are found in, e.g., \cite{SLM_MIMO}, but those required considerable computations, cost, and complexity. Model-driven approaches to simultaneously reduce $\mathsf{PAPR}$ over all antennas were also proposed. In \cite{SLM_3}, instead of applying $\mathsf{SLM}$ to each antenna, the sequence with the highest $\mathsf{PAPR}$ over all transmit antennas was selected. Usually, $\mathsf{SLM}$ and $\mathsf{PTS}$ methods demand side information (SI) to be sent to the receiver along with each transmitted data block for retrieving the original data. The need for SI requires extra bandwidth overhead, and the incorrect detection of the SI bits over the channel will lead to significant degradation in the $\mathsf{BER}$ performance of the receiver in the $\mathsf{MIMO}$-$\mathsf{OFDM}$ system. 
\vspace{-0.5cm}
\subsection{Deep-Learning-Based Schemes (Data Driven) for PAPR Reduction}
\vspace{-0.3cm}
In recent years much research has been dedicated to applying $\mathsf{DL}$ techniques in the design and optimization of wireless communication networks, e.g., \cite{modelorai, ai4b5g, NN_detector}. Several papers have proposed $\mathsf{DL}$ methods to handle $\mathsf{PAPR}$ reduction. For example, the authors of \cite{PAPR_Sohn,PAPR_Sohn_Kim}, added a $\mathsf{NN}$ to reduce the complexity of the active constellation scheme, followed by $\mathsf{CF}$.
In \cite{cshape3,AE_DCO} the authors present an $\mathsf{AE}$ solution for $\mathsf{PAPR}$ reduction, while minimizing the $\mathsf{BER}$ degradation.
In \cite{PAPR_yara} a $\mathsf{CAE}$ was suggested for the implementation of an end-to-end $\mathsf{SISO-OFDM}$ communication network that simultaneously reduces the $\mathsf{PAPR}$ and reconstructs the transmitted symbols, while keeping acceptable spectral requirements. Another learning-based approach, which considers the reduction of the $\mathsf{PAPR}$ and $\mathsf{ACPR}$ together with the maximization of the achievable information rate for a single-carrier waveform above multipath channels, was proposed in \cite{waveformLearning}.    
The authors in \cite{AE_SLM_ACO} proposed a deep $\mathsf{NN}$ combined with $\mathsf{SLM}$ to mitigate the high $\mathsf{PAPR}$ issue of $\mathsf{OFDM}$ signal types. 

All of the above papers consider a $\mathsf{SISO}$ network. A $\mathsf{PAPR}$ reduction scheme assisted by $\mathsf{DL}$ for a $\mathsf{MIMO}$-$\mathsf{OFDM}$ system was suggested in \cite{ML_PAPR_MIMO}. The authors apply selective tone reservation \cite{str_mimo} on each antenna separately and then apply unused beam reservation \cite{ubr_mimo} on all antennas together. An $\mathsf{ML}$-based method for approximating the optimal tabular hyperparameters required for using selective tone reservation and unused beam reservation was suggested.
\vspace{-0.4cm}
\subsection{Main Contributions}\vspace{-0.3cm}
Some of the aforementioned $\mathsf{PAPR}$ reduction approaches suffer from in-band interference, out-of-band distortions, and high computational complexity. Moreover, published $\mathsf{ML}$-based solutions mostly handle single antenna scenarios. Those who deal with $\mathsf{PAPR}$ reduction for $\mathsf{MIMO}$ systems use $\mathsf{ML}$ only for the $\mathsf{PAPR}$ reduction block and not for the end-to-end network implementation.  
This paper aims to handle the $\mathsf{PAPR}$ problem in $\mathsf{MIMO}$ systems as an integral part of a waveform design objective. In particular, we design a communication system that simultaneously achieves $\mathsf{PAPR}$ reduction, acceptable spectral behavior of the PA's output, and good $\mathsf{BER}$ performance. The suggested end-to-end network aims to resolve the $\mathsf{MIMO}$ detection problem as a part of the other mentioned objectives. To the best of our knowledge, this approach is new.
Novelties we introduce include using a $\mathsf{CAE}$ combined with a gradual loss learning technique to handle the multi-objective optimization of the network, and adding the effect of the $\mathsf{HPA}$ on an integrated end-to-end $\mathsf{MIMO}$ communication system. 
We present an iterative $\mathsf{MIMO}$ detection algorithm integrated into transmitter-receiver end-to-end communication system joint optimization.
We demonstrate our algorithm's results on 5G $\mathsf{MIMO}$-$\mathsf{OFDM}$ Matlab toolbox simulated data, and we compare our method with classical methods for $\mathsf{PAPR}$ reduction and waveform design, and show competitive results for all the objectives mentioned above. The proposed algorithm offers performance improvement for future wireless communication systems. 
We show that our model provides competitive $\mathsf{PAPR}$ reduction, waveform design, and detection results.

The rest of this paper is structured as follows.  In Section II, the problem is defined and formulated separately for $\mathsf{MIMO}$ detection, and for $\mathsf{PAPR}$ reduction as a part of the $\mathsf{MIMO}$-$\mathsf{OFDM}$ system. We then present the proposed DL-based system architecture for the multi-objective optimization and explain the training procedure in Section III. Section IV provides detailed numerical simulation results and insights. Finally, Section V gives concluding remarks.
\vspace{-0.4cm}
\section{Notation and Problem Definition}
In this section, we introduce the notation and the problem definition.
\vspace{-0.6cm}
\subsection{Notation}
\vspace{-0.3cm}
Throughout this paper, we use the following notations. The set of real numbers is denoted by $\mathbb{R}$, while the set of complex numbers is denoted by $\mathbb{C}$. Random variables will be denoted by capital letters, and their realizations will be denoted by lower-case letters, e.g., $X$ and $x$, respectively. Calligraphic letters denote sets, e.g.,  $\mathcal{X}$. We use the notation $X^n$ to denote the random vector $(X_1,X_2,\dots,X_n)$ and $x^n$ to denote the realization of such a random vector. The expectation operator is denoted by $\mathbb{E}\left[\cdot\right]$. $(\cdot)^{*}$, $(\cdot)^{\dag}$ denote the conjugate, and pseudo-inverse operators, respectively.
\vspace{-0.4cm}
\subsection{Problem Definition}
\vspace{-0.2cm}
In this section, we describe mathematically each part of the integrated problem of $\mathsf{MIMO}$ detection together with $\mathsf{PAPR}$ reduction and spectrum constraints. 
First, we give a brief introduction to the end-to-end setup used in our system. 

\subsubsection{MIMO detection model}\label{sssec:mimo_model}
Let us assume a standard $\mathsf{MIMO}$-$\mathsf{OFDM}$ system with $N_t$ transmit antennas and $N_r$ receive antennas. Transmission is considered over a memoryless complex-valued channel model, while assuming frequency flatness and slow fading. A $\mathsf{MIMO}$ system can be modeled by the following complex baseband model:
\vspace{-0.5cm}
\begin{align} \label{eq:mimo_model}
\boldsymbol{y} = \boldsymbol{H}\boldsymbol{x} + \boldsymbol{n},
\vspace{-0.5cm}
\end{align}
where $\boldsymbol{x}\in\mathbb{C}^{N_t}$ is the transmitted complex symbol vector drawn from a finite discrete constellation of size $|\mathcal{M}|$, $\boldsymbol{H}\in\mathbb{C}^{N_r\times N_t}$ is the complex baseband channel matrix that is related to a specific subcarrier, 
$\boldsymbol{n}\in\mathbb{C}^{N_r}$ is complex background $\mathsf{AWGN}$ seen at the receiver, and $\boldsymbol{y}\in\mathbb{C}^{N_r}$ is the received complex vector resulting from the propagation of the transmitted symbols through the channel contaminated by $\mathsf{AWGN}$.

As the proposed implementation is based on a real-valued $\mathsf{NN}$ model determined by the $\mathsf{DL}$ Pytorch library, \eqref{eq:mimo_model} is expressed with real values by splitting and concatenating each signal into its real and imaginary parts:
 \begin{align}   \label{eq:reparam_sig}
    \boldsymbol{x}=\begin{bmatrix}
        Re\{\boldsymbol{x}\}\\
        Im\{\boldsymbol{x}\}\\
    \end{bmatrix},\ 
    \boldsymbol{y}=\begin{bmatrix}
        Re\{\boldsymbol{y}\}\\
        Im\{\boldsymbol{y}\}\\
    \end{bmatrix}, \ 
    \boldsymbol{n}=\begin{bmatrix}
        Re\{\boldsymbol{n}\}\\
        Im\{\boldsymbol{n}\}\\
    \end{bmatrix},
     \boldsymbol{H}=\begin{bmatrix}
        Re\{\boldsymbol{H}\} & -Im\{\boldsymbol{H}\}\\
        Im\{\boldsymbol{H}\} & Re\{\boldsymbol{H}\}\\
    \end{bmatrix}.
 \end{align}

In the $\mathsf{MIMO}$ detection problem, the objective is to detect the transmitted symbols, $\boldsymbol{x}$, given the received data $\boldsymbol{y}$.
The optimal solution for the $\mathsf{MIMO}$ detection of the transmitted symbols problem defined above is given by the $\mathsf{MLE}$ algorithm, that is,
\begin{align} \label{ml_solution}
\hat{\boldsymbol{x}}_{mle} = \argminA_{\boldsymbol{x}\in\mathcal{X}^{N_t}} ||\boldsymbol{y}-\boldsymbol{H}\boldsymbol{x}||^2,
\end{align}
where $\mathcal{X}$ denotes the set of possible transmitted symbols (i.e., signal constellation). The solution of \eqref{ml_solution} requires an exhaustive search over all $|\mathcal{M}|^{N_t}$ possible transmitted vectors. Therefore, it is infeasible for an actual implementation where large-scale $\mathsf{MIMO}$ setups and/or a large constellation are in use. 

\subsubsection{PAPR problem in MIMO-OFDM}\label{sssec:papr_mimo}
In an OFDM system with ${N}$ complex orthogonal subcarriers, the discrete-time
transmitted $\mathsf{OFDM}$ signal at the $n_t$ antenna, is given by the inverse discrete Fourier transform ($\mathsf{IDFT}$):
\begin{align}
    {x}_{n_t,n} = \frac{1}{\sqrt{{N}}}\sum_{k=0}^{{N}-1}{X}_{n_t,k}e^{j\frac{2\pi}{{L}{N}}kn}, \;\;  0\le n\le {L}{N}-1, \;\;  1\le n_{t}\le N_{t},\label{eqn:model}
\end{align}
where $\{{X}_{n_t,k}\}_{k=0}^{{N}-1}$ are random input symbols per antenna, modulated by a finite constellation, and ${L}\geq1$ is the over-sampling factor (${L}=1$ is the Nyquist sampling rate).
As shown in \cite{PAPR_OVER_2}, oversampling by a factor of four results in a good approximation of the continuous-time $\mathsf{PAPR}$ of complex $\mathsf{OFDM}$ signals. 
The discussed problem considers non-linear $\mathsf{HPA}$s at each of the $N_t$ TX branches. We assume that the $\mathsf{HPA}$s in all branches have the same non-linear characteristic, which is a reasonable assumption, considering current wireless $\mathsf{MIMO}$ systems. Also, in a discrete implementation, the same $\mathsf{HPA}$s are usually used. 

The $\mathsf{PAPR}$ of the transmitted signal in \eqref{eqn:model} is defined as the ratio between the maximum peak power and the average power of the $\mathsf{OFDM}$ signal. Specifically, the $\mathsf{PAPR}$ at the $n_t$-th transmit antenna is defined by:
\begin{align}
\mathsf{PAPR_{n_t}} \triangleq \frac{\max_{0 \le n \le {LN}-1} |{x_{n_t,n}}|^2}{\mathbb{E}|{x_{n_t,n}}|^2}.
\end{align}
For the entire $\mathsf{MIMO}$-$\mathsf{OFDM}$ system, the $\mathsf{PAPR}$ reduction method we use will consider the maximum $\mathsf{PAPR}$ among all $N_t$ transmit antennas, as the same PA model is used in all branches:
\begin{align}
\mathsf{PAPR_{MIMO-OFDM}} = \max_{1 \le n_t \le N_t} {\mathsf{PAPR_{n_t}}}.
\end{align}

As $\mathsf{HPA}$ non-linearity causes spectral regrowth, an important assessment for the spectral purity of the system is the $\mathsf{ACPR}$ criterion, which is the ratio between the power of the adjacent channel and the power of the main channel. Following \cite{3gpp}, we define it as
\begin{align}
\mathsf{ACPR} \triangleq \frac{\max{\left(\int_{\s{BW}/2}^{3\s{BW}/2} \s{P_{ss}}(f)\;\mathrm{d}f ,\int_{-3\s{BW}/2}^{\s{BW}/2} \s{P_{ss}}(f)\;\mathrm{d}f\right)}}{\int_{-\s{BW}/2}^{\s{BW}/2}\s{P_{ss}}(f)\;\mathrm{d}f},
\end{align}
where $\s{P_{ss}}(\cdot)$ is the power spectral density ($\mathsf{PSD}$) of the signal at the $\mathsf{HPA}$'s output, and $\s{BW}$ is the primary channel bandwidth, which is assumed to be equal to the data signal bandwidth.

\begin{wrapfigure}{r}{0.6\textwidth}
\vspace{-1.1cm}
  \centering
    \begin{psfrags}
    \psfragscanon
        \psfrag{A}[][][0.8]{$X_k$}
        \psfrag{B}[][][0.9]{$x_n$}
        \psfrag{C}[][][0.8]{PAPR \& Spectral}
        \psfrag{D}[][][0.8]{Optimization}
        \psfrag{E}[][][1]{Enc}
        \psfrag{F}[][][0.9]{Filter}
        \psfrag{G}[][][0.9]{${x}_n^\s{F}$}
        \psfrag{H}[][][0.8]{Power}
        \psfrag{I}[][][0.8]{Amplifier}
        \psfrag{J}[][][1]{$\mathrm{G}\left[\cdot\right]$}
        \psfrag{K}[][][0.9]{${x}_n^\s{P}$}
        \psfrag{L}[][][0.9]{Channel}
        \psfrag{M}[][][0.7]{+}
        \psfrag{N}[][][0.9]{$y_n$}
        \psfrag{O}[][][1]{DFT}
        \psfrag{P}[][][0.8]{$Y_k$}
        \psfrag{R}[][][0.9]{Reconstruct}
        \psfrag{S}[][][0.8]{$\hat{{X}}_k$}
        \psfrag{T}[][][1]{IDFT}
        \psfrag{V}[][][0.9]{\& Detect}
        \psfrag{W}[][][1]{$\mathbf{w}_n$}
    \psfragscanoff
    \includegraphics[scale = 0.83]{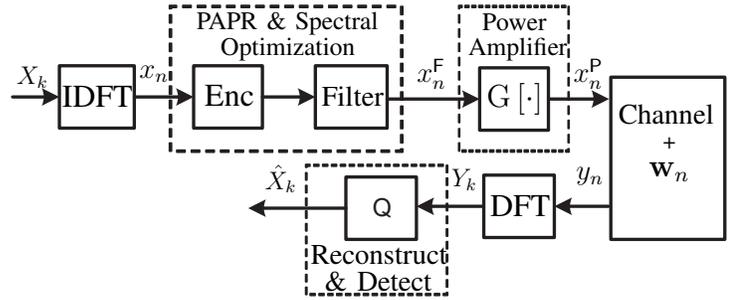}
    \caption{General system model diagram.\vspace{-0.9cm}}
    \label{fig:block_model}
\end{psfrags}
\end{wrapfigure}

A block diagram of the communication system model is shown in Fig.~\ref{fig:block_model}. 
Specifically, the encoder and filter blocks mitigate the $\mathsf{PAPR}$ effect and design the waveform to comply with predefined spectral mask requirements. For example, the encoder block can model a clipping operation, while the filter can be a standard band-pass filter ($\mathsf{BPF}$). 
The filtered signal ${x}_n^\s{F}$ is amplified by a non-linear $\mathsf{HPA}$.
The amplified signal, ${x}_n^\s{P}=\mathrm{G}({x}_n^\s{F})$, is transmitted through a fading channel with $\mathsf{AWGN}$.
The channel decoder receives the noisy signal and attempts to reconstruct and detect the transmitted signal. For model-driven approaches, a classical detection algorithm, e.g., $\mathsf{MLE}$, is applied for detecting the estimated symbol denoted by $\hat{{X}}_k$. 

The role of the $\mathsf{HPA}$ is to convert the low-level transmission
signal to a high power signal, capable of driving the antenna at the desired power level.
The $\mathsf{HPA}$ has to operate close to its saturation region for maximal power efficiency. If the $\mathsf{HPA}$ exceeds the saturation point and enters the non-linear area of operation, the output signal becomes non-linear.
Accordingly, to operate the amplifier only in the linear region, we need to make sure that the amplifier operates at a power level that is lower than the saturation point. This is achieved by down-scaling the input signal by an input back-off ($\mathsf{IBO}$) factor. The drawback of adding the $\mathsf{IBO}$ attenuation is that the output power decreases, which makes the $\mathsf{HPA}$ power-inefficient. 

\begin{wrapfigure}{r}{0.45\textwidth}
  \centering
    \psfrag{M}[][][0.8]{$A_{\mathsf{in}}$}
    \psfrag{Q}[][][0.8]{$A_{\mathsf{out}}$}
    \psfrag{Y}[][][0.6]{$3dB$}
    \includegraphics[width = 7.1cm]{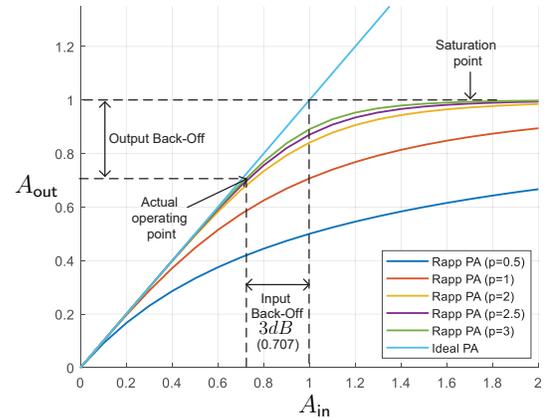}
    \vspace{-0.1cm}
    \caption{RAPP HPA output versus input signal for different smoothness $p$ values.}
    \label{fig:RAPP_out}
\end{wrapfigure}
There are several commonly used models for the non-linearity of an $\mathsf{HPA}$. Here, we will focus on the RAPP behavioral amplifier model \cite{rapp_1}, which is very accurate for solid-state-power amplifiers. The model's AM/AM conversion is given by 
\begin{align}
    \s{G}({A_{\mathsf{in}}}) = v\cdot{A_{\mathsf{in}}}\cdot\left(1+\left(\frac{v{A}_{\mathsf{in}}}{{A}_0}\right)^{2p}\right)^{-\frac{1}{2p}\vspace{-0.9cm}},
\end{align}
where ${A_{in}}$ is the input amplitude, ${A}_0$ is the limiting output amplitude, $v$ is the small signal gain,
${p}$ is a smoothness parameter controlling the transition from the linear region to the saturation region, 
and $\s{G}({A})$ is the output amplitude. Figure \ref{fig:RAPP_out} shows RAPP $\mathsf{HPA}$ outputs versus input for several smoothing factor values.
\section{Proposed Waveform Design Structure}
\begin{figure}[t]
     \centering
     \begin{subfigure}[b]{1\textwidth}
         \centering
            \psfrag{A}[][][0.7]{$X_k$}
            \psfrag{B}[][][0.7]{Zero}
            \psfrag{C}[][][0.7]{pad}
            \psfrag{D}[][][0.7]{IFFT}
            \psfrag{E}[][][0.9]{$x_n$}
            \psfrag{F}[][][0.8]{2}
            \psfrag{G}[][][0.7]{calc}
            \psfrag{M}[][][0.9]{\;\;$x^B_n$}
            \psfrag{N}[][][0.7]{Filter}
            \psfrag{O}[][][0.7]{BO}
            \psfrag{P}[][][0.7]{PA}
            \psfrag{Q}[][][0.9]{\;\;$x^F_n$}
            \psfrag{R}[][][0.6]{PAPR}
            \psfrag{S}[][][0.6]{Calculation}
            \psfrag{T}[][][0.6]{$ \mathcal{L}_{2b}$}
            \psfrag{U}[][][0.9]{$x^P_n$}
            \psfrag{V}[][][0.7]{MIMO}
            \psfrag{W}[][][0.7]{Channel}
            \psfrag{X}[][][0.8]{$X_k$}
            \psfrag{Y}[][][0.9]{$y_n$}
            \psfrag{Z}[][][0.6]{ACPR}
            \psfrag{a}[][][0.6]{$ \mathcal{L}_3$}
            \psfrag{b}[][][0.6]{Loss}
            \psfrag{c}[][][0.6]{$ \mathcal{L}_1$}
            \psfrag{d}[][][0.6]{Reconstruction}
            \psfrag{e}[][][0.67]{Encoder $f(x)$}
            \psfrag{f}[][][0.8]{$\alpha$}
            \psfrag{g}[][][0.67]{Decoder $g(y)$}
            \psfrag{h}[][][0.6]{$N_{t}\times{N_{sc}}$}
            \psfrag{i}[][][0.55]{$N_{t}$}
            \psfrag{j}[][][0.55]{$N_{r}$}
            \psfrag{k}[][][0.7]{$Y_k$}
            \psfrag{m}[][][0.9]{\;\;$x^E_n$}
            \psfrag{n}[][][0.7]{unpad}
            \psfrag{p}[][][0.6]{$N_{r}\times{N_{sc}}$}
            \psfrag{q}[][][0.7]{FFT}
            \psfrag{t}[][][0.9]{Transmitter}
            \psfrag{u}[][][0.9]{Receiver}
            \psfrag{x}[][][0.7]{$\hat{{X}}_k$}
            \psfrag{w}[][][0.9]{$+ w_n$}
            \psfrag{y}[][][0.6]{$ \mathcal{L}_{2a}$}
            \includegraphics[scale = 0.67]{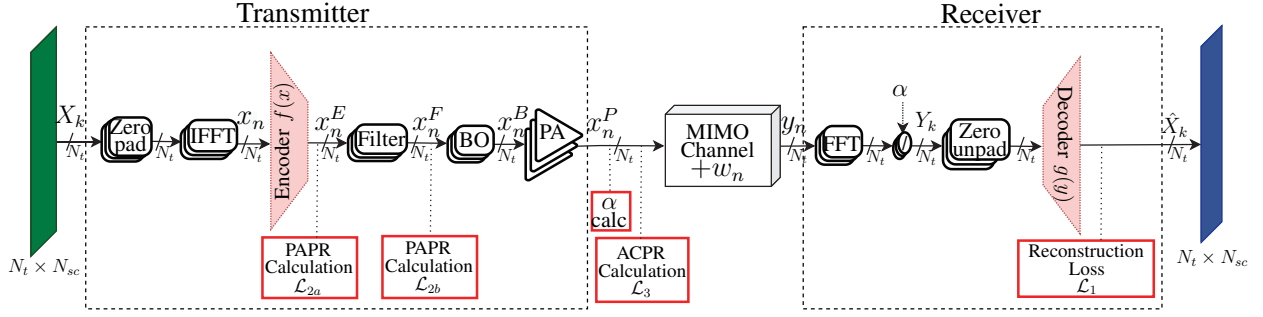}
            \caption{Conv-AE overall scheme.\vspace{0.4cm}}
            \label{fig:NN_scheme}
     \end{subfigure}
     \hfill
     \begin{subfigure}[b]{1\textwidth}
         \centering
            \psfrag{A}[][][0.7]{Input Tensor}
            \psfrag{B}[][][0.7]{Conv+BN+GELU}
            \psfrag{C}[][][0.7]{Fully Connected}
            \psfrag{D}[][][0.7]{Predicted Tensor}
            \psfrag{E}[][][0.7]{Kernel}
            \psfrag{N}[][][0.7]{$N_{sc}$}
            \psfrag{P}[][][0.7]{Power Norm Layer}
            \psfrag{T}[][][0.7]{$N_t$}
            \psfrag{U}[][][0.8]{ch2}
            \psfrag{a}[][][0.7]{$ \mathcal{L}_2$}
            \psfrag{b}[][][0.7]{loss}
            \psfrag{c}[][][0.7]{$ \mathcal{L}_3$}
            \psfrag{e}[][][0.7]{fc1}
            \psfrag{f}[][][0.8]{conv1}
            \psfrag{g}[][][0.8]{conv2}
            \psfrag{h}[][][0.9]{PAPR Reduction Block - Encoder $f(x)$}
            \psfrag{i}[][][0.8]{conv3}
            \psfrag{t}[][][0.6]{3}
            \psfrag{u}[][][0.8]{ch1}
            \psfrag{v}[][][0.6]{1}
            \includegraphics[scale = 0.67]{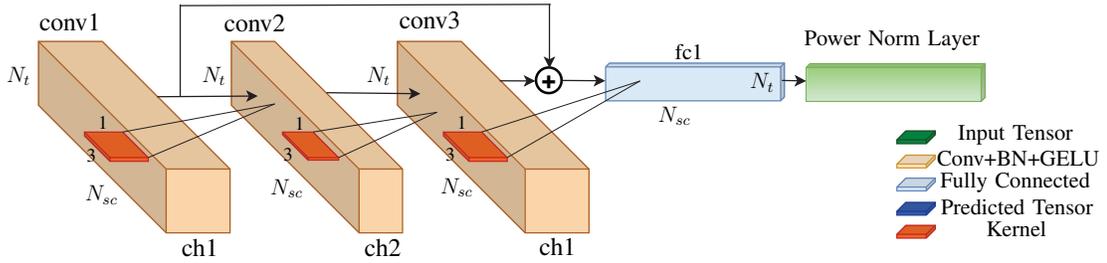}
            \caption{PAPR reduction block - Encoder scheme $f(x)$.\vspace{0.4cm}}
            \label{fig:Enc_scheme}
     \end{subfigure}
     \hfill
     \begin{subfigure}[b]{1\textwidth}
         \centering
            \psfrag{A}[][][0.52]{$X_{i-1}$}
            \psfrag{B}[][][0.52]{$H^{*}HX_{k+1}$}
            \psfrag{C}[][][0.52]{$X_{k+1}$}
            \psfrag{H}[][][0.52]{$H^{*}H()$}
            \psfrag{L}[][][0.8]{$i_{th}$ iteration}
            \psfrag{N}[][][0.7]{$N_{sc}$}
            \psfrag{T}[][][0.7]{$N_r$}
            \psfrag{U}[][][0.7]{ch2}
            \psfrag{a}[][][0.52]{$H^{*}Y$}
            \psfrag{b}[][][0.52]{$H^{*}HX_{k}$}
            \psfrag{c}[][][0.52]{$X_{k}$}
            \psfrag{d}[][][0.52]{$H^{*}HX_{i-1}$}
            \psfrag{e}[][][0.6]{Concat}
            \psfrag{f}[][][0.65]{conv4}
            \psfrag{g}[][][0.65]{conv5}
            \psfrag{h}[][][1]{Detection Block - Decoder $g(x)$}
            \psfrag{i}[][][0.6]{fc2 + Softmax}
            \psfrag{j}[][][0.6]{Layer}
            \psfrag{k}[][][0.8]{$k_{th}$ iteration}
            \psfrag{t}[][][0.55]{$3$}
            \psfrag{u}[][][0.7]{ch1}
            \psfrag{v}[][][0.7]{fc2}
            \psfrag{x}[][][0.6]{$X_i$}
            \includegraphics[scale = 0.44]{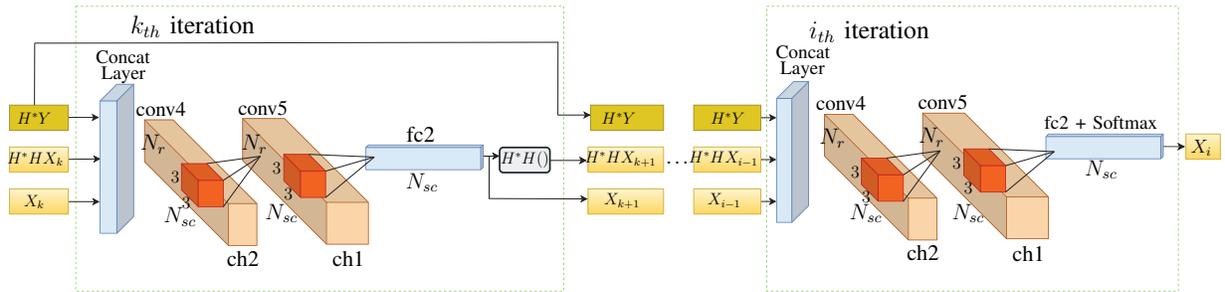}
            \caption{Detection block - Decoder scheme $g(x)$.\vspace{-0.4cm}}
            \label{fig:Dec_scheme}
     \end{subfigure}
        \caption{Structure of the proposed conv-AE.\vspace{-1.3cm}}
        \label{fig:three structure graphs}
\end{figure}

In this section, we describe our multi-objective optimization $\mathsf{CAE}$ model architecture. Motivated by research evidence of powerful learning ability, under the same conditions of the $\mathsf{MIMO}$-$\mathsf{OFDM}$ examined structure, it is expected that the proposed $\mathsf{CAE}$ model will achieve good enough performance to be compared with classical $\mathsf{PAPR}$ reduction methods combined with the $\mathsf{MLE}$ detector. 
We first briefly discuss the general $\mathsf{CAE}$ concept.
Then, we describe our algorithm building blocks, and the joint $\mathsf{PAPR}$ reduction, spectral design, and detection operation in detail.
The motivation and structure of the iteration-based decoder with regard to handling the $\mathsf{MIMO}$ detection problem as a part of the $\mathsf{CAE}$ network will be explained. 
The proposed architecture in Fig. \ref{fig:three structure graphs} is then elaborated, including the Bussgang's non-linearity compensation theorem, followed by a description of the gradual learning process. Last, the training procedure of the $\mathsf{CAE}$ multi-objective optimization network operation will be described.

Fig. \ref{fig:NN_scheme} shows the overall end-to-end communication network structure implemented by a $\mathsf{CAE}$ model, where the encoder and the decoder are the trainable blocks. Each mentioned loss component is calculated using the operations appearing in the red blocks in the figure. It can be noticed that there are two $\mathsf{PAPR}$ calculation-based loss components, as will be detailed later, that helped achieve improved spectral behavior and $\mathsf{BER}$ results. Fig. \ref{fig:Enc_scheme} describes the encoder structure. It is constructed of 2D convolutional layers to be compatible with the input example dimensions, a fully connected layer, and a power normalization layer at the output. We also added a skip connection to improve stability and performance. Fig. \ref{fig:Dec_scheme} shows the decoder structure. It is also implemented with 2D convolutional layers. To be able to optimize the network under different communication channel realizations, we used an iterative-based solution to the MIMO detection problem.
We emphasize that the presented $\mathsf{CAE}$ model, once trained on a single training set, presents competitive results for the discussed multi-objective optimization problem in the presence of a wide range of noise power levels, without any knowledge of the SNR level. 
\vspace{-0.7cm} 
\subsection{Convolutional Autoencoder (CAE)}
\vspace{-0.2cm} 
The proposed implementation uses an $\mathsf{AE}$ learning system based on a convolutional neural network ($\mathsf{CNN}$). 
The general structure of an $\mathsf{AE}$ consists of two main blocks: the encoder $f(\f{x})$ and the decoder $g(\f{x})$, where $\f{x}$ is the input data. The $\mathsf{AE}$ is trained to minimize a certain joint loss function, which we denote by $\mathcal{L}(\f{x},g(f(\f{x})))$.
An end-to-end communication system can be interpreted as an $\mathsf{AE}$ in which the encoder and the decoder are part of the transmitter and the receiver, respectively, and can jointly optimize them through an end-to-end learning procedure. 

$\mathsf{CNN}$s are widely used for feature extraction and pattern recognition in $\mathsf{ML}$ models. Compared with a fully connected ($\mathsf{FC}$) network, a CNN has significantly fewer connections between adjacent layers, and thus fewer parameters and weights to train, resulting in lower complexity and much faster training. 
\vspace{-0.6cm} 
\subsection{Iteration-Based Model Motivation}
\vspace{-0.2cm} 
An iterative approach based on convolutional layers, was adopted to implement the decoder. The $\mathsf{MIM0}$ detector was designed to solve the $\mathsf{MLE}$ optimization (\ref{ml_solution}) using the projected gradient descent approach, where the input to the detector is a linear combination of features instead of the channel output directly. Utilizing such detectors significantly improved the detection results when various channel realizations were integrated. 
The detector input feature selection was based on the following projected gradient decent approach used to optimize (\ref{proj_sgd}):
\vspace{-0.4cm}
\begin{align} \label{proj_sgd}
\vspace{-0.4cm}
    ||\boldsymbol{y}-H\boldsymbol{x}||^2.
    \vspace{-0.7cm}
\end{align}
As shown in \cite{det_net, NN_detector}, such iterations are updated by
\begin{align}
    \hat{\boldsymbol{x}}_{k+1} &= P_c\left[\hat{\boldsymbol{x}}_{k}-\delta_{k}\frac{\partial||\boldsymbol{y}-\boldsymbol{H}\boldsymbol{x}||^2}{\partial\boldsymbol{x}}\biggr\rvert_{\mathbf{\boldsymbol{x}=\boldsymbol{\hat{x}}_k}}\right]= P_c\left[\hat{\boldsymbol{x}}_{k}-\delta_{k}\left(\boldsymbol{H}^H\boldsymbol{y} - \boldsymbol{H}^H\boldsymbol{H}\hat{\boldsymbol{x}}_k\right)\right], 
\end{align}
where $\hat{\boldsymbol{x}}_{k}$ is the objective estimation in the $k$'th iteration, $\boldsymbol{H}^H$ indicates the conjugate transpose operation over the channel matrix $\boldsymbol{H}$, $\delta_k$ is the step size, and $P_c$ is the non-linear projection operator. The above can be adapted to a deep $\mathsf{NN}$ iterative solution by the following linear combination form:
\vspace{-0.25cm}
\begin{align}   \label{iter_solution}
\vspace{-0.2cm}
    \hat{\boldsymbol{x}}_{k+1} = P_c(\hat{\boldsymbol{x}}_k + \delta_{1k} \boldsymbol{H}^H\boldsymbol{y} + \delta_{2k} \boldsymbol{H}^H\boldsymbol{H}\hat{\boldsymbol{x}}_k),
    \vspace{-0.2cm}
\end{align}
where $\delta_{1k}$ and $\delta_{2k}$ are learned hyper-parameters to be optimized.
One of the motivations for using the described iterative decoder approach was a former work published in \cite{det_net, NN_detector}. In our work, it is designed as part of the joint encoder-decoder multi-task optimization. Also, it was implemented with convolutional layers, that enabled better computational and performance capabilities for the joint optimization end-to-end system, and the per subcarrier/antenna alternating analysis. A softmax layer is added at the output of the $\mathsf{CAE}$ to generate probabilistic outputs.
In the following section, we provide the complete detection procedure.
\vspace{-0.45cm} 
\subsection{Proposed CAE Architecture}
This section introduces the suggested $\mathsf{CAE}$ learning network implementation of the $\mathsf{MIMO}$-$\mathsf{OFDM}$ system for the multi-objective optimization task. 
We consider a $\mathsf{MIMO}$-$\mathsf{OFDM}$ scheme with $\mathsf{N_t}$ transmit antennas and $\mathsf{N_r}$ receive antennas, where the $\mathsf{OFDM}$ is of order $\mathsf{K}$. The input is represented by a matrix in the frequency domain, i.e.
\begin{align}
\underline{\mathbf{X}} = 
\begin{pmatrix}
X^{(1)}(1) & X^{(1)}(2) & \cdots & X^{(1)}(K) \\
X^{(2)}(1) & X^{(2)}(2) & \cdots & X^{(2)}(K) \\
\vdots  & \vdots  & \ddots & \vdots  \\
X^{(N_t)}(1) & X^{(N_t)}(2) & \cdots & X^{(N_t)}(K)
\end{pmatrix},
\label{eq:tx_mat}
\end{align}
where, for any $\mathsf{n_t}\in{[1, N_t]}$ and $k\in{[1, K]}$, $X^{(n_t)}(k)$ is a M-QAM constellation complex-valued symbol.

In Fig. \ref{fig:NN_scheme}, we illustrate the general structure of the end-to-end communication network implemented by the CAE configuration. Specifically, we consider a transmitter that takes the two-dimensional matrix $\underline{\mathbf{X}}$ as an input. The transmitter's output goes through a $\mathsf{MIMO}$ channel, together with $\mathsf{AWGN}$. Finally, the noisy channel outputs are fed into a receiver to estimate $\underline{\mathbf{X}}$. The operations within the transmitter and the receiver are described below.
\vspace{-0.1cm}
\begin{itemize}
    \item \textbf{Transmitter:} the input signal $\underline{\mathbf{X}}$ is zero-padded on the subcarriers' dimension and converted to the time domain via an $\mathsf{IFFT}$ applied on each of the transmitter branches, outputting $\{{x}_n\}_{n=0}^{{{LN}-1}}$. These symbols serve as the input to the encoder, which acts as a $\mathsf{PAPR}$ reduction block, followed by a $\mathsf{BPF}$ filter for optimizing the spectral behavior by reducing the out-of-band radiation. Its frequency response is a rectangular window with the same bandwidth as $X^{(n_t)}_k$. Then, a predefined $\mathsf{IBO}$ is applied just before the signal is amplified by the $\mathsf{HPA}$.
    \item \textbf{Receiver:} the distorted $\mathsf{OFDM}$ symbols are divided by an $\alpha$ factor to compensate for the non-linear distortions, as will be detailed in the following. Finally, the proposed $\mathsf{CAE}$ decoder reconstructs and detects the estimated $\mathsf{MIMO}$-$\mathsf{OFDM}$ transmitted signals.
\end{itemize}
\vspace{-0.1cm}
The encoder comprises three convolutional layers, and the decoder is composed of iterative construction of convolutional layers. Each convolutional layer is followed by a non-linear activation function and batch normalization \cite{batch_norm}, and then a fully connected layer. In addition, a residual connection is added to the encoder block, which sums (element-wise) the input to the second convolutional layer and the output of the third convolutional layer. It turns out that this modification improves the overall performance of the suggested scheme significantly. The intuition is that adding another path for data to reach the latter parts of the $\mathsf{NN}$ makes it easier to optimize the mapping \cite{res_deep}. Furthermore, the encoder has a power normalization layer, which ensures that the transmitted signal meets the power constraints of unit average energy per $\mathsf{OFDM}$ symbol. This way, the intended SNR is maintained.
We tested several activation functions, including sigmoid, rectified linear unit ($\mathsf{RELU}$), Gaussian error linear unit ($\mathsf{GELU}$), and scaled exponential linear unit ($\mathsf{SELU}$) \cite{SELU}. Empirically, it was found that $\mathsf{SELU}$ activation provides the best results for our $\mathsf{CAE}$ scheme.

As illustrated in Fig. \ref{fig:NN_scheme} and Fig. \ref{fig:Enc_scheme}, since the encoder is responsible for the $\mathsf{PAPR}$ reduction, which is calculated per $\mathsf{OFDM}$ symbol, we start with per antenna analysis, where each antenna is treated separately. A one-dimensional kernel per TX branch handles this. 
The encoder architecture can be described by the following:
\begin{align}   \label{iter_solution_param}
   f(\boldsymbol{x}) &= \rho_{L_{f}}\left(\biggr\rvert\boldsymbol{W}^{f}_{L_{f}}\left(\rho_{L_{f}-1}\left(...\left(\rho_{1}\left(\biggr\rvert\boldsymbol{W}^{f}_{1}\boldsymbol{x}+\boldsymbol{b}^{f}_{1}\biggr\rvert_{bnorm}\right)\right)\right)...\right)\right.\nonumber\\
   &\left.\ \ \ \ \ \ \ \ \ +\boldsymbol{b}^{f}_{L_{f}}\biggr\rvert_{bnorm}+\left(\biggr\rvert\boldsymbol{W}^{f}_{1}\boldsymbol{x}+\boldsymbol{b}^{f}_{1}\biggr\rvert_{bnorm}\right)\right),
\end{align}
where $L_f$ is the number of the encoder's convolutional layers, $\boldsymbol{W}_{i}^{f}$, and $\boldsymbol{b}_{i}^{f}$ are the encoder's weight matrix and bias vector, respectively, for the $i$'th layer, with size determined as a part of the network design. $\rho_{i}(\cdot)$ is the activation function of the $i$'th layer, and $bnorm$ means the layer passes through a batch normalization.

The next part of this process applies the non-linear $\mathsf{HPA}$s on each TX branch of the transmitter time domain signals, each composed of all subcarriers. 
The signal is then converted via $\mathsf{FFT}$ to the frequency domain, and the zero-unpadding block removes the out-of-band samples. Afterward, frequency domain analysis is performed on each subcarrier transmitted through all $N_t$ antennas. Each subcarrier is transmitted through its related complex baseband channel described by a $\left(N_{r} \times N_{t}\right)$ matrix, and the $\mathsf{AWGN}$ is added as well. 

To continue with the per subcarrier analysis on the receiver side, we need to overcome the non-linearity of the $\mathsf{HPA}$. To that end, we compensate the receiver input signal by applying an attenuation factor represented by $\alpha$. Bussgang's decomposition theorem \cite{Bussgang} states that if a zero-mean Gaussian signal passes through a memoryless non-linear device, then the output-input cross-correlation function is proportional to the input autocovariance. Accordingly, the value of $\alpha$ is chosen to minimize the variance of the non-linear signal distortions, such that it is attempted that the transmitted signal in each transmitter branch is linearly separated, and thus represented as a sum of the signal and distortion. It can be shown that
\vspace{-0.25cm}
\begin{align}
    \mathsf{\alpha} = \frac{\mathbb{E}\left({x_n^\s{F}}{\overline{x}_n^\s{P}}\right)}{\mathbb{E}\left(|{x_n^\s{F}}|^2\right)}, 
\end{align}
\vspace{-0.25cm}
where ${{x}_n^\s{P}}$ is the complex output signal of the PA, and ${\overline{x}_n^\s{P}}$ is its complex conjugate.
By assuming that the $\mathsf{PSD}$ of the in-band distortion is approximately flat \cite{psd_flat} in the frequency domain, the output signal of the $\mathsf{HPA}$ on the $k$-th subcarrier can then be expressed as
\vspace{-0.25cm} 
\begin{align}
    X^\s{P}(k) = \alpha(k)X^\s{F}(k)+D(k),
\end{align}
\vspace{-0.25cm} 
where $D(k)$ is the non-linear distortion on the $k$-th subcarrier.
The same model is assumed for all $\mathsf{PA}$s; therefore, it can be concluded that $\mathsf{\alpha_{n_t}} = \mathsf{\alpha}$.

At the $\mathsf{MIMO}$ decoder, Fig. \ref{fig:Dec_scheme}, the per subcarrier analysis is continued, meaning that different subcarriers of the same RX branch will not be mixed. To generalize our end-to-end structure, we modified the 1D kernel to a 2D kernel at the decoder part. It also helped better reconstruct the signal after the encoder layers. As explained in the previous sub-section, we use an iterative procedure to implement the decoder, which is designed for signal reconstruction and detection. A general mathematical description of one iteration, $k$, of the presented decoder is given by
\begin{align}   
    \begin{split}
    \boldsymbol{d}_k &= \begin{pmatrix}\boldsymbol{\hat{x}}_{k-1}, \delta_{1k} \boldsymbol{H}^H\boldsymbol{y}, \delta_{2k} \boldsymbol{H}^H\boldsymbol{H}\hat{\boldsymbol{x}}_{k-1}\end{pmatrix}
    \\\boldsymbol{\hat{x}}_{k} &= g(\boldsymbol{d}_k) \\&= \rho_{L_{g},k}\left(\biggr\rvert\boldsymbol{W}^{g}_{L_{g},k}\left(\rho_{L_{g}-1,k}\left(...\left(\rho_{1,k}\left(\biggr\rvert\boldsymbol{W}^{g}_{1,k}\boldsymbol{d}_{k}^{T}+\boldsymbol{b}^{g}_{1,k}\biggr\rvert_{bnorm}\right)\right)\right)...\right)+\boldsymbol{b}^{g}_{L_{g},k}\biggr\rvert_{bnorm}\right),\nonumber
    \end{split}
\end{align}
where $L_g$, $\boldsymbol{W}_{i}^{g}$, and $\boldsymbol{b}_{i}^{g}$, have the same definitions as described for the encoder's block, only that these apply for the decoder. The decoder's input features vector, $\boldsymbol{d}_k$, was initialized by randomizing a prediction $\hat{\boldsymbol{x}}_{0}$. Initialization by zeros resulted in performance degradation. 
\vspace{-0.6cm}
\subsection{Training of the CAE Network}
\vspace{-0.25cm}
We train a single $\mathsf{CAE}$ model for all tested $\mathsf {SNR}$ values.
We use the AdamW optimizer \cite{adamw} that runs back-propagation to optimize the model during training. This optimizer is designed to improve gradients when $\text{L}_2$ regularization is used. Our loss function is set to solve the constrained optimization problem by handling three objectives: accurate signal reconstruction (minimal $\mathsf{BER}$), minimal $\mathsf{PAPR}$, and acceptable $\mathsf{ACPR}$.  

We solve this constrained optimization problem by recasting it as an unconstrained problem by constructing the Lagrangian function and augmenting the objective function with a quadratic penalty term \cite{boyd2011distributed}. The augmented Lagrangian ($\mathsf{AL}$) combines the Lagrangian formulation with a weighted quadratic penalty function. The general $\mathsf{AL}$ for an inequality-constrained problem can be described by
\begin{align}
\vspace{-0.4cm}
    \mathcal{F}_{\rho^k}(\mathbf{x},\lambda^k) = f(\mathbf{x})+\lambda_1^{k}c_1(\mathbf{x})+\frac{1}{2}\rho_1^{k}\norm{c_1(\mathbf{x})}_2^2+\frac{1}{2\rho_2^{k}}\left\{\left[\max \{0,\lambda_2^{k}+\rho_2^kc_2(\mathbf{x})\}\right]^2-(\lambda_2^k)^2\right\},
    \label{eq:aug_tot_loss_mod}
\end{align}
\vspace{-0.2cm}
where $f$ denotes the objective function, $\rho^k\triangleq(\rho_1^k,\rho_2^k)$ are positive penalty parameters, $\lambda^k\triangleq(\lambda_1^k,\lambda_2^k)$ are the Lagrangian multipliers, the $c_1$-involved expressions handle the equality constraint, and $c_2$ is for the inequality constraint. Equation \eqref{eq:aug_tot_loss_mod} considers the elimination of a slack variable $s\geq0$ that was introduced in the representation of the inequality constraint to transform it into a relaxed equality constraint. As suggested in \cite{Constrained_optimization}, the minimizer $s=\max \{0, c_2(x)-\lambda_{2}\frac{1}{\rho_{2}}\}$ was used. $k$ is the iteration number for updating the Lagrangian multipliers and penalty term, according to the following rule derived by the dual ascent method \cite{boyd2011distributed},
\begin{align}
    x^{k+1} &:= \argminA_{x}\mathcal{F}_{\rho^k}(\mathbf{x},\lambda^k)
    \vspace{-0.4cm}
    \label{eq:loss}\\
    \lambda_1^{k+1} &:= \lambda_1^{k}+\rho_1^{k}c_{1}(\mathbf{x}^{k+1})
    \vspace{-0.2cm}
    \label{eq:lambda_aug}\\
    \lambda_2^{k+1} &:= \max \{0,\lambda_2^{k}+\rho_2^{k}c_{2}(\mathbf{x}^{k+1})\}.
    \vspace{-0.2cm}
    \label{eq:lambda_slack}
\end{align}

\vspace{-0.3cm}
We saw better convergence and more stable results for different BO values by adding the quadratic penalty function and adaptively updating the multipliers instead of keeping them constant. Since adaptive penalty parameter update was not beneficial for the examined cases, it was added as a fixed hyperparameter.

Following the above-described general inequality constraint optimization problem, the formulation of our loss function based on the appropriate objective and constraints, represented by four loss components $\mathcal{L}_1$, $\mathcal{L}_{2a}$, $\mathcal{L}_{2b}$, and $\mathcal{L}_3$, is
\begin{align} \label{eq:loss_total}
\vspace{-0.3cm}
    \mathcal{L}(\bold{x},\hat{\bold{x}}, \lambda_{2a}^k, \lambda_{2b}^k,  \lambda_3^k) &= \mathcal{L}_1 (\bold{x},\hat{\bold{x}})+\lambda_{2a}^{k}\mathcal{L}_{2a} (\bold{x})+\frac{\rho_{2a}}{2}\norm{\mathcal{L}_{2a}(\bold{x})}_2^2   \nonumber\\ &+ \lambda_{2b}^{k}\mathcal{L}_{2b} (\bold{x})+\frac{\rho_{2b}}{2}\norm{\mathcal{L}_{2b}(\bold{x})}_2^2 
       +\frac{1}{2\rho_3}\left\{\left[\max \{0,\lambda_3^{k}+\rho_3\mathcal{L}_3(\bold{x})\}\right]^2-(\lambda_3^k)^2\right\},
\end{align}
where $\lambda_{2a}$, $\lambda_{2b}$, $\lambda_3$, $\rho_{2a}$, $\rho_{2b}$, and $\rho_3$ are the appropriate Lagrange multipliers and penalty parameters, accordingly. These are considered hyper-parameters, which balance the contribution of each loss component to the joint loss function. 
We start with a moderate value of $\lambda_{2a}$, $\lambda_{2b}$ and $\lambda_3$ and then iterate for a better value according to the resulting $\mathsf{PAPR}$ loss of each iteration and some predetermined $\mathsf{PAPR}$ threshold value. Better performance was observed for relatively small $\lambda_{2b}$ values, with very low $\rho_{2b}$, meaning that  $\lambda_{2b}$ was kept almost constant during training.      

The loss function we use for optimizing the signal reconstruction and detection is the sum of negative log loss function of the predicted output probability of the real and imaginary parts of each symbol, with $\text{L}_2$ regularization to reduce over-fitting. Denoting by $x$ the input sample (which is also the output target), $\hat{x}$ as the estimated signal, $\Theta$ as the model's weights, and $\lambda_1$ as a hyperparameter for tuning the $\text{L}_2$ regularization, the loss function for each $\mathsf{OFDM}$ $\mathsf{MIMO}$ sample is given by,
\begin{align} \label{loss}
    \mathcal{L}_1 (\bold{x},\hat{\bold{x}}) &= -\Bigg[ 
    \sum_{j=1}^{N_t}
    \sum_{s=1}^{N_{sc}}
    \sum_{q=1}^{N_c}1 \Big\{ Re \{x^j\} = l_q\Big\}
    \log P_{\theta}\Big(Re\left\{\hat{x}^j\right\}= l_q   \Big)  \nonumber \\ &+ 
       \sum_{j=1}^{N_t}
       \sum_{s=1}^{N_{sc}}
    \sum_{q=1}^{N_c}1 \Big\{ Im \{x^j\} = l_q\Big\}
    \log P_{\theta}\Big(Im\left\{\hat{x}^j\right\} = l_q   \Big)
    \Bigg] + \lambda_1 \norm{\Theta}_2^2,
\end{align}
where $N_c = \sqrt{|\mathcal{M}|}$ denotes the number of the real value possibilities, $l_q$, of each of the real and imaginary parts of the transmitted modulated symbol.  

The $\mathsf{PAPR}$ minimization part is handled with two loss components, where one, $\mathcal{L}_{2a}$, is calculated according to the $\mathsf{BPF}$ input, ${{x}_n^\s{E}}$, and the other one, $\mathcal{L}_{2b}$, according to the $\mathsf{BPF}$ output, ${{x}_n^\s{F}}$ (cf. Fig. \ref{fig:NN_scheme}). These are our equality constraints, defined by
\begin{align}
    \mathcal{L}_{2a} (x) & = \mathsf{PAPR}\{{{x}_n^\s{E}}\}, \\ 
    \mathcal{L}_{2b} (x) & = \mathsf{PAPR}\{{{x}_n^\s{F}}\}.  
\end{align}
Other than the role of $\mathsf{PAPR}$ minimization handled by either of the components, $\mathcal{L}_{2a} (x)$, significantly improved the $\mathsf{BER}$ result together with the $\mathsf{ACPR}$, while $\mathcal{L}_{2b} (x)$, enabled us to control and obtain better spectral performance, meaning lower $\mathsf{ACPR}$ with lower output back-off ($\mathsf{OBO}$) values.
The $\mathsf{ACPR}$ loss component is given by
\begin{align}
    \mathcal{L}_3 (x) & = \mathsf{ACPR}\{{{x}_n^\s{P}}\}-\mathsf{ACPR_{req}},  
\end{align}
where ${{x}_n^\s{P}}$ is the PA's output, and $\mathsf{ACPR_{req}}$ is the required $\mathsf{ACPR}$ value, which is usually dictated by a standard. 
$\mathsf{ACPR_{req}}$ was set according to the 5G standard requirements for high spectral purity: $\mathsf{ACPR_{req}}\leq{-45}\mathrm{dB}$ \cite{3gpp}; thus, $\mathcal{L}_3 (x)$ defines our inequality constraint. 

We have applied a gradual loss learning technique.
In the first stage, the loss function consisted only of $\mathcal{L}_1$ and optimized only the reconstruction loss. Then, after a predetermined number of epochs, the loss function defined in \eqref{eq:loss_total} was used to reduce the $\mathsf{PAPR}$ and improve the spectral behavior. The gradual loss learning enables better control and stability in tuning the different criteria' trade-offs.   
\vspace{-0.7cm}
\section{Results and Insights}
\subsection{Data Generation and Experimental Setup}
To train and test the proposed data-driven model, the MATLAB\textsuperscript{\textregistered} 5G Toolbox\textsuperscript{\texttrademark} \cite{Matlab5G} was used. This toolbox provides 5G radio-standard-compliant functions to generate accurate data for $\mathsf{MIMO}$-$\mathsf{OFDM}$ transmission, according to specified constellation sizes and examined $\mathsf{MIMO}$ setups.
$\mathsf{MIMO}$-$\mathsf{OFDM}$ transmissions over fading channels were simulated, where TDL-D type channels - a 13 delay tap channel with a 30ns delay spread, as described in the 3gpp specification document \cite{3gpp}, were used for our implemented algorithm.

We consider a $\mathsf{MIMO}$-$\mathsf{OFDM}$ system with $K = $72 subcarriers over 14 $\mathsf{OFDM}$ symbols per frame. 4375 batches of 32 $\mathsf{MIMO}$ samples each were used for a single training set, where the input and output of the $\mathsf{CAE}$ sample shape is $[2/4-antennas, {(72-subcarriers)}\times{(2-complex-parts)}\times{(4-oversampling)}]$. An oversampling factor $L=4$, and smoothness factor $p=2$ were considered.
We trained three identical $\mathsf{CAE}$ models on the following setups:
\begin{enumerate}
\vspace{-0.2cm}
    \item QPSK modulation scheme with a $2\times2$ $\mathsf{MIMO}$ setting, with 3GPP multipath channel.
    \item 16-QAM modulation scheme with a $4\times4$ $\mathsf{MIMO}$ setting, with 3GPP multipath channel.
    \item 16-QAM modulation scheme with a $4\times4$ $\mathsf{MIMO}$ setting, with AWGN channel.
    \vspace{-0.2cm}
\end{enumerate}
To provide an unbiased performance evaluation of the final training model, the training and test data sets were generated independently, i.e. $\mathsf{OFDM}$ symbols, channel realization, and noise were randomized independently.
In the following, we give numerical performance results of our multi-objective $\mathsf{CAE}$ model compared to a classical $\mathsf{CF}$ algorithm with a clipping ratio of 4.08 dB, and to $\mathsf{SLM}$ with $U=64$ phase sequences, with $\mathsf{MLE}$ added for $\mathsf{MIMO}$ detection. The inference part was performed on 7000 $\mathsf{MIMO}$ samples for each SNR point.
\vspace{-0.5cm}
\subsection{Training Setup}
\vspace{-0.2cm}
As a part of the experimental analysis, we performed an extensive exploration of different model structures and hyper-parameters, including the number of layers, kernel sizes, number of convolutional layer channels, regularization, dropout, number of decoder iterations, batch-normalization, learning rate, $\mathsf{AL}$ parameters, trained SNR value, and epoch number. We found that the best performance versus complexity on both examined $\mathsf{MIMO}$ setups was achieved for the same model structure, only with different training data sets.
As the constellation, the number of subcarriers per OFDM symbols, and the number of antennas are higher, the training is longer, and it is harder to achieve the desired results.
\begin{table}[t]
    \begin{center}
    \caption{CAE Proposed Structure}
\begin{tabular}{c||c|c|c|c||c|c|c|c}
     & \multicolumn{4}{c||}{\textbf{Transmitter}} & \multicolumn{4}{c}{\textbf{Receiver}}  \\\hline
     Parameter&Value&Kernel&Ch-in&Ch-out&Value&Kernel&Ch-in&Ch-out  \\\hline\hline
     Input size & $4\times720$ &-&-&-&$12\times144$&-&-&\\\hline
     Conv (SELU)& - & $1\times3$ & 1 & 21 & - & $3\times3$ & 1 & 15 \\\hline
     Conv (SELU)& - & $1\times3$ & 21 & 15 & - & $3\times3$ & 15 & 21 \\\hline
     Conv (SELU)& - & $1\times3$ & 15 & 21 & - & - & - & - \\\hline
     FC (Linear) output size & $4\times720$  &-&-&-& $12\times144$&-&-&-  \\\hline 
     Decoder iterations&-&-&-&-& 10&-&-&- \\\hline
\end{tabular}
\bigskip
\bigskip
\label{CAE_Struct}
\begin{tabular}{c|c|c|c|c|c|c}
    Conv padding&LR&Epochs num&Grad start&SNR train&$\lambda_{2a}^{(0)}$, $\lambda_{2b}^{(0)}$, $\lambda_3^{(0)}$&$\rho_{2a}$, $\rho_{2b}$, $\rho_{3}$  \\\hline\hline
    2& 0.001 & 140 & 45 & 40 dB & 0.015, 0.001, 0.005 & 0.0015, 0.00001, 0.001 \vspace{-0.9cm}
\end{tabular}
\vspace{-0.6cm}
\label{CAE_def}
\end{center}
\end{table}
The proposed $\mathsf{CAE}$ structure for the above $4\times4$ $\mathsf{MIMO}$ system is described in Table \ref{CAE_Struct}, where $\lambda_{2a}^{(0)}$, $\lambda_{2b}^{(0)}$, and $\lambda_3^{(0)}$ are the values of the first iteration when the $\mathsf{AL}$ epochs start, and 'Grad start' indicates the number of initial epochs where only the reconstruction loss is counted, optimizing the unconstrained problem. 'LR' indicates the learning rate. Training on any of the data sets with the same best SNR value used for noise generation, 'SNR train', showed the top overall inference performance for any tested SNRs. Adding dropout had no benefit in all examined setups. 
\vspace{-0.7cm}
\subsection{BER Analysis}
\begin{figure}[H]
\vspace{-0.6cm}
     \centering
     \begin{subfigure}[b]{0.49\textwidth} 
         \centering
         \includegraphics[width=\textwidth]{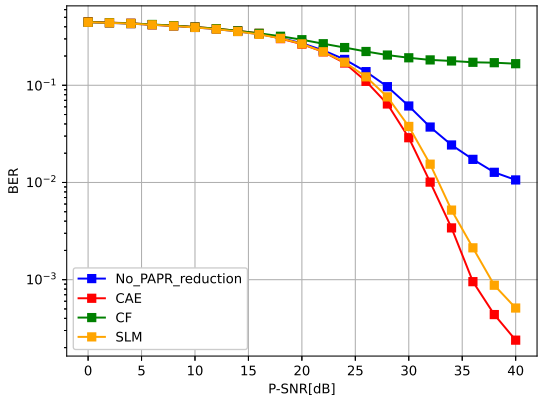}
         \caption{16-QAM, 4x4 MIMO with 3GPP multipath channel}
         \label{fig:BER_testing_16QAM}
     \end{subfigure}
     \hfill
     \begin{subfigure}[b]{0.49\textwidth}
         \centering
         \includegraphics[width=\textwidth]{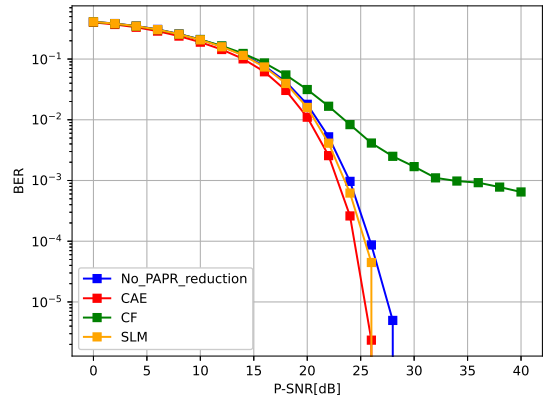}
         \caption{16-QAM, 4x4 MIMO with AWGN channel}
         \label{fig:BER_testing_AWGN_16QAM}
     \end{subfigure}
     \hfill
     \begin{subfigure}[b]{0.49\textwidth}
     \vspace{-0.2cm}
         \centering
         \includegraphics[width=\textwidth]{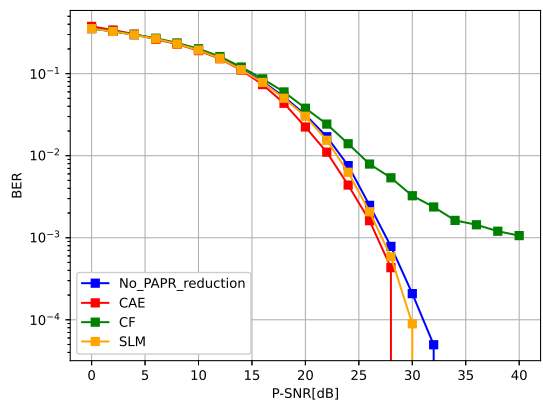}
         \caption{4-QAM, 2x2 MIMO with 3GPP multipath channel}
         \label{fig:BER_testing_4QAM}
     \end{subfigure}
        \caption{BER vs. $\s{P\_SNR}$ of the considered methods and setups.\vspace{-0.85cm}}
        \label{fig:three ber graphs}
\end{figure}

\vspace{-0.3cm}
The calculation of $\mathsf{BER}$ versus Peak Signal to Noise Ratio ($\s{P\_SNR}$) is used here as a key parameter to measure the reconstruction and detection of the transmitted signal. Considering a normalized channel, i.e. $\mathbb{E}\norm{\bold{H}}^2=1$, the $\s{P\_SNR}$ is defined as the ratio between the $\mathsf{MIMO}$ system maximal emitted energy, $P_T$, and the noise power, $\sigma_w^2$, such that
\begin{align}
\vspace{-0.4cm}
    \s{P\_SNR} = \frac{P_T}{\sigma_w^2}.
    \vspace{-0.4cm}
\end{align}
\vspace{-0.1cm}
As shown in Fig. \ref{fig:three ber graphs}, the $\mathsf{CAE}$ has competitive $\mathsf{BER}$ vs. $\s{P\_SNR}$ performance compared to the other standard examined methods in most of the $\s{P\_SNR}$ range, where a visible gain is achieved at the higher part. As the NN does not assume any specific physical model, it has better robustness to distortions. That is, the $\mathsf{MIMO}$-$\mathsf{OFDM}$ signal reconstruction and detection of $\mathsf{HPA}$-distorted data as a part of the multi-objective optimization proposed by our end-to-end $\mathsf{DL}$ scheme has the benefit over the common algorithms.
\vspace{-0.65cm}
\subsection{CCDF for PAPR Comparison}
\begin{figure}[H]
\vspace{-0.2cm}
     \centering
     \centering
     \begin{subfigure}[b]{0.49\textwidth}
         \centering
         \includegraphics[width=\textwidth]{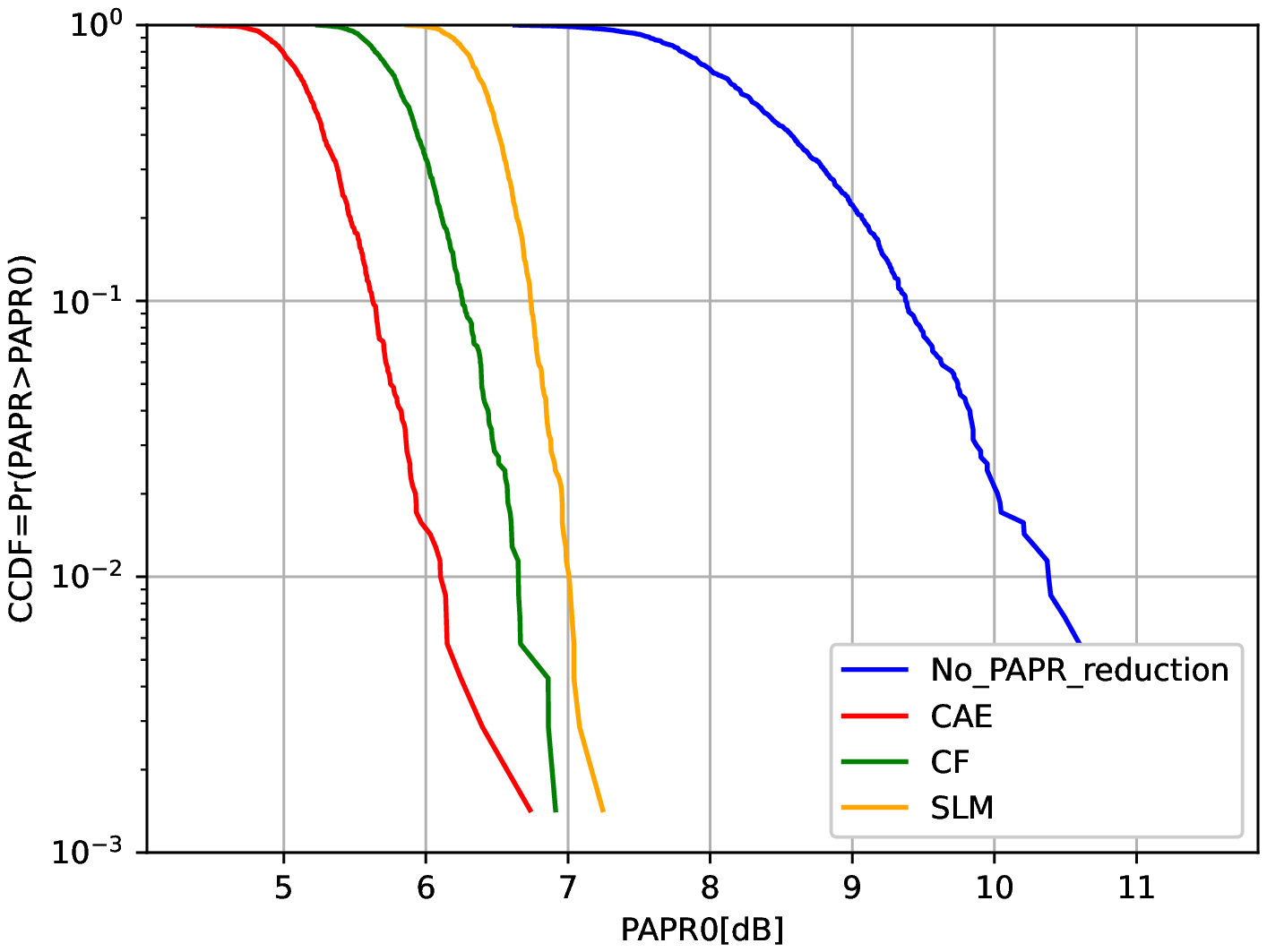}
         \caption{16-QAM, 4x4 MIMO with 3GPP multipath channel}
         \label{fig:PAPR_testing_16QAM}
     \end{subfigure}
     \hfill
     \begin{subfigure}[b]{0.49\textwidth}
         \centering
         \includegraphics[width=\textwidth]{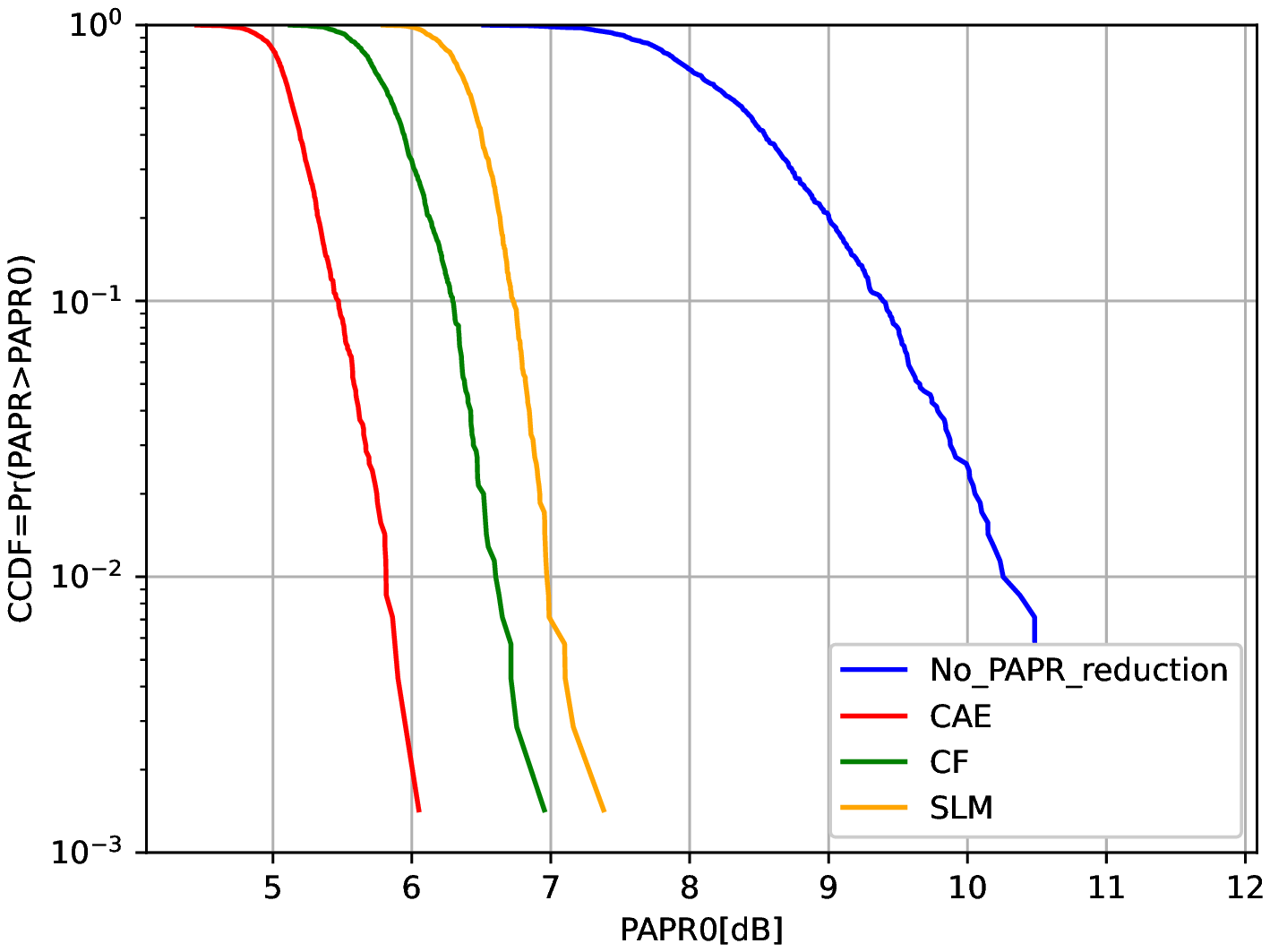}
         \caption{16-QAM, 4x4 MIMO with AWGN channel}
         \label{fig:PAPR_9_16QAM_AWGN}
     \end{subfigure}
     \hfill
     \begin{subfigure}[b]{0.49\textwidth}
         \centering
         \includegraphics[width=\textwidth]{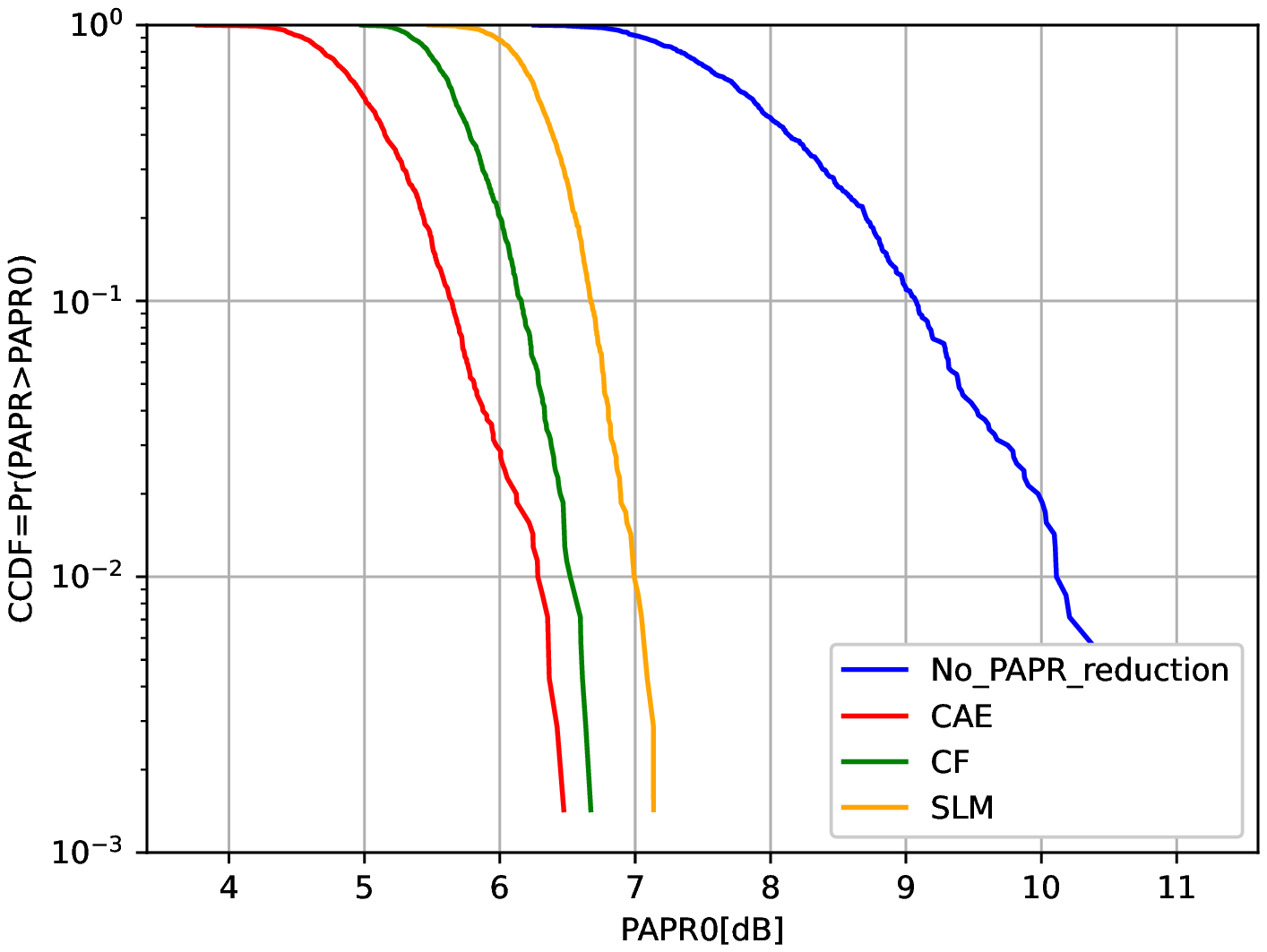}
         \caption{4-QAM, 2x2 MIMO with 3GPP multipath channel}
         \label{fig:PAPR_testing_4QAM}
     \end{subfigure}
        \caption{CCDF of PAPR of the considered methods.\vspace{-0.4cm}}
        \label{fig:three ccdf graphs}
\end{figure}

To demonstrate the $\mathsf{PAPR}$ performance of different methods, complementary cumulative distribution function (CCDF) curves are presented in Fig. \ref{fig:three ccdf graphs}. The CCDF of the $\mathsf{PAPR}$ denotes the probability that the $\mathsf{PAPR}$ exceeds a certain threshold, i.e. $\mathbb{P}(\mathsf{PAPR_{MIMO-OFDM}} > \mathsf{PAPR_{0}})$. The $\mathsf{PAPR}$ is calculated according to the $\mathsf{BPF}$ output, ${x}_n^\s{F}$. 
As can be observed in Fig. \ref{fig:three ccdf graphs}, the proposed $\mathsf{CAE}$ achieves the better performance of $\mathsf{PAPR}$ reduction compared to the $\mathsf{CF}$ and $\mathsf{SLM}$ methods. However, still the $\mathsf{BER}$ and spectral behavior are more important for performance evaluation.
\vspace{-1.2cm}
\subsection{Spectrum Analysis}
\begin{figure}[t]
\vspace{-0.4cm}
     \centering
     \begin{subfigure}[b]{0.494\textwidth}
         \centering
         \includegraphics[width=\textwidth]{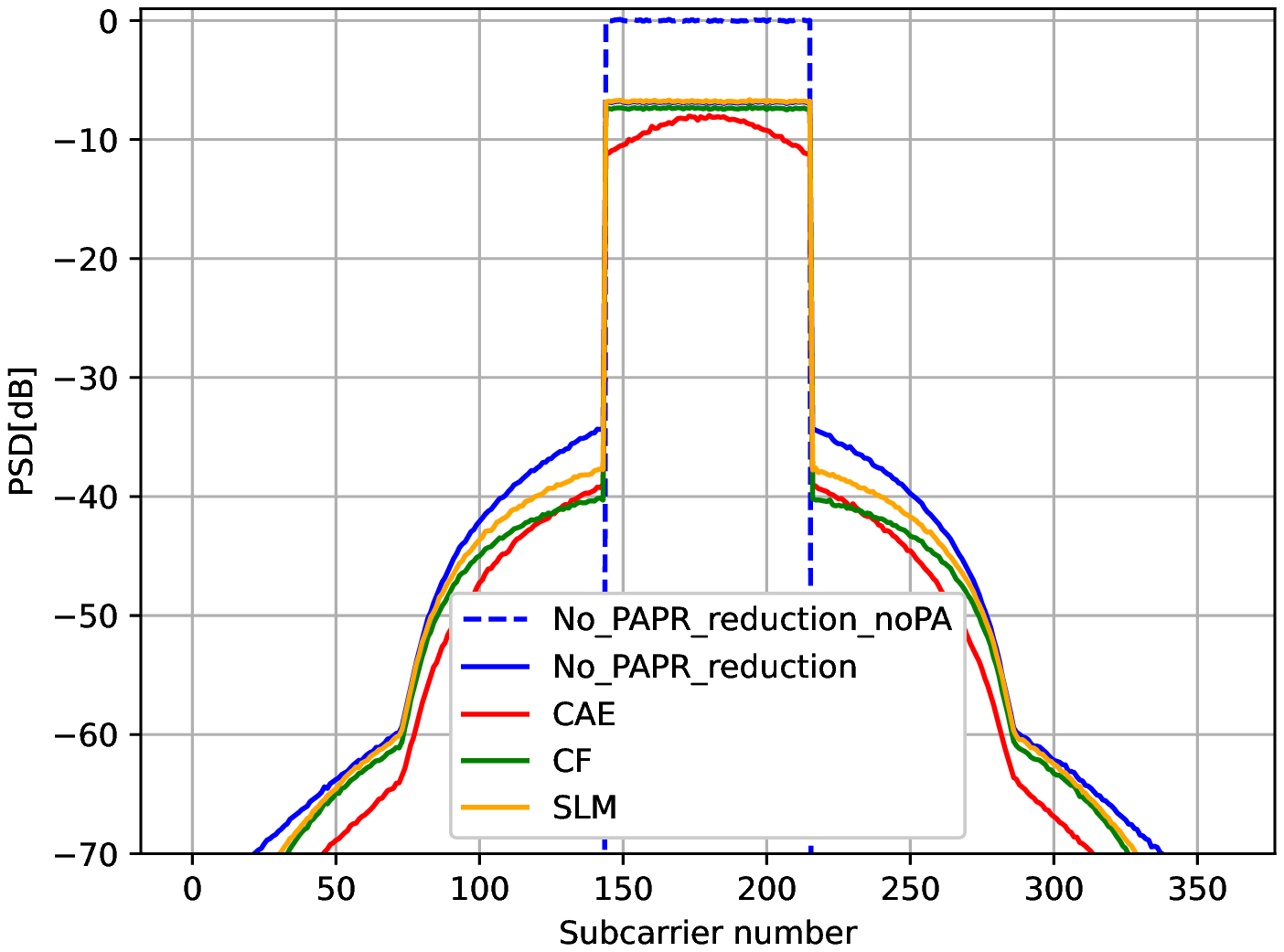}
         \caption{16-QAM, 4x4 MIMO with 3GPP multipath channel}
         \label{fig:spectral_mask_16QAM}
     \end{subfigure}
     \hfill
     \begin{subfigure}[b]{0.494\textwidth}
         \centering
         \includegraphics[width=\textwidth]{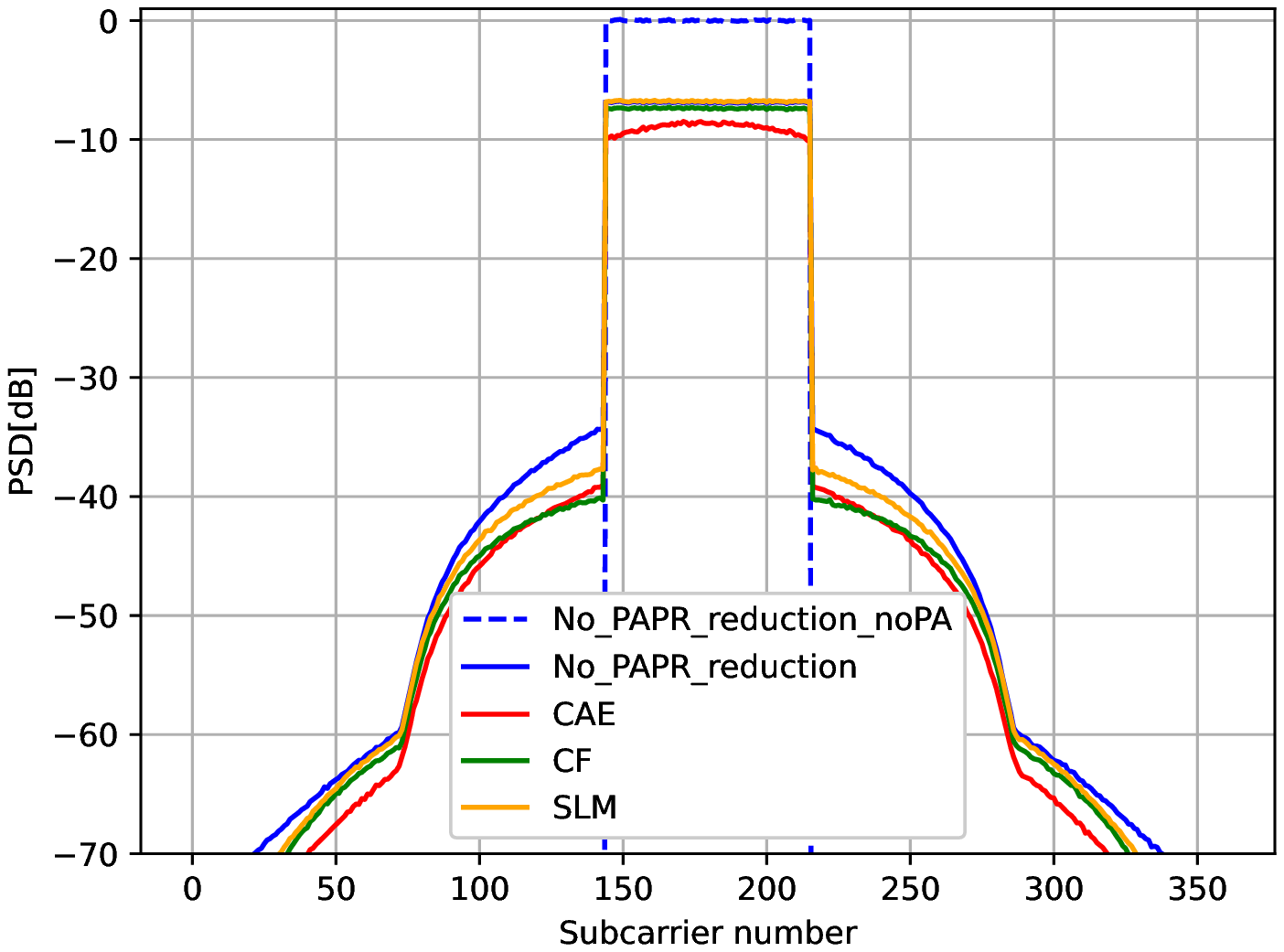}
         \caption{16-QAM, 4x4 MIMO with AWGN channel}
         \label{fig:spectral_mask_AWGN_16QAM}
     \end{subfigure}
     \hfill
     \begin{subfigure}[b]{0.49\textwidth}
         \centering
         \includegraphics[width=\textwidth]{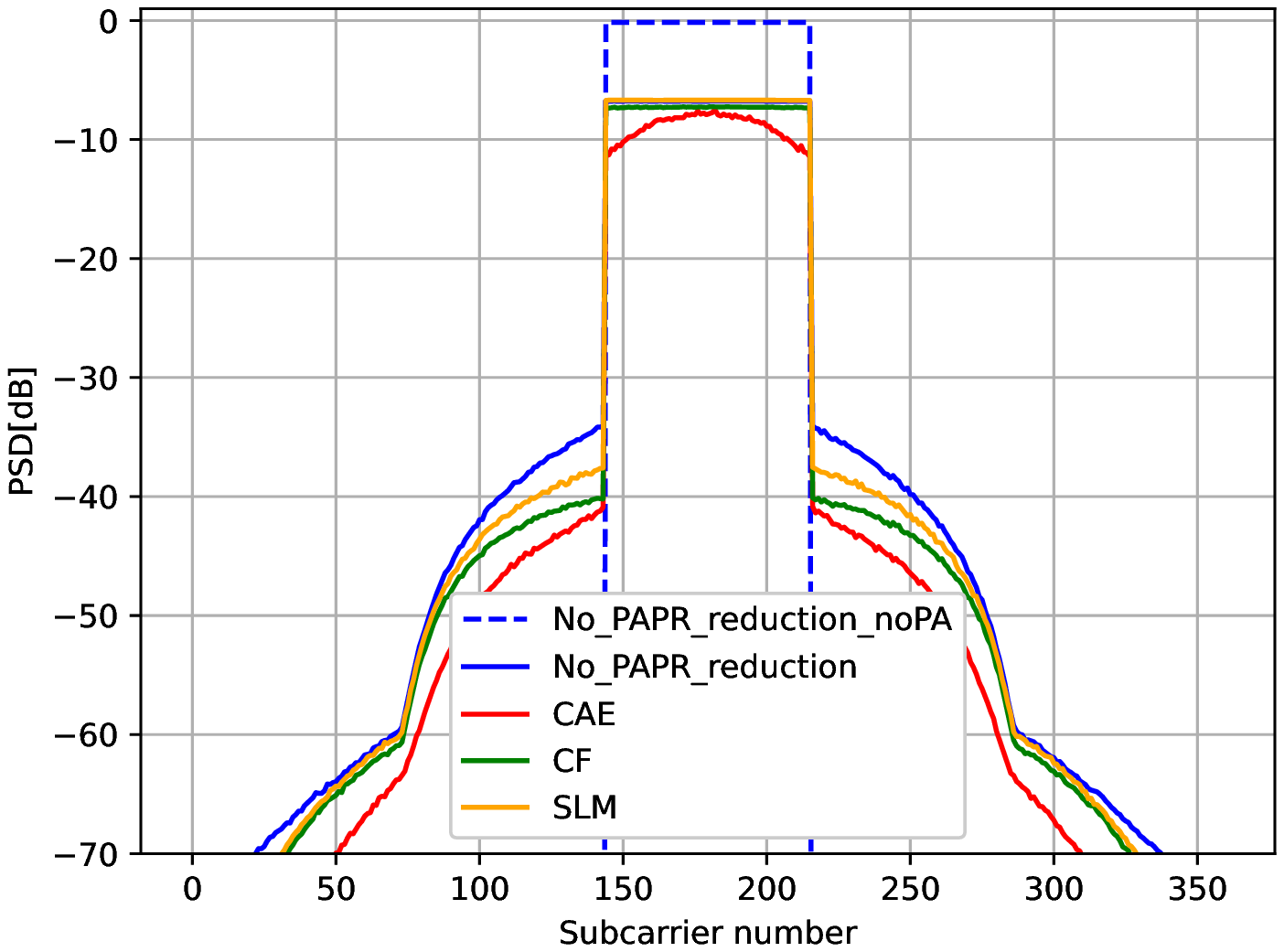}
         \caption{4-QAM, 2x2 MIMO with 3GPP multipath channel}
         \label{fig:spectral_mask_4QAM}
     \end{subfigure}
        \vspace{-0.5cm}
        \caption{PSD for the considered methods.\vspace{-1.2cm}}
        \label{fig:three spectrum graphs}
\end{figure}

Figure \ref{fig:three spectrum graphs} compare the spectral performance in terms of the $\mathsf{PSD}$ of the transmitted signals for all examined methods. The dashed rectangle shows perfect spectral behavior for a linear $\mathsf{HPA}$ with no non-linear components. 

\begin{figure}[H]
\centering
\vspace{-0.3cm}
\begin{tabular}{cc}
\subfloat[moderate $\lambda_{2b}$]{\includegraphics[scale = 1.1]{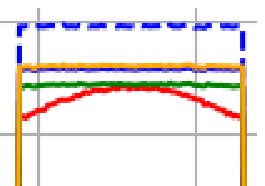}} 
   \subfloat[high $\lambda_{2b}$]{\includegraphics[scale = 1.1]{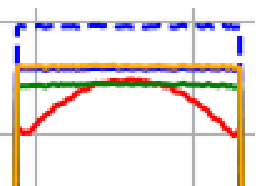}}
\subfloat[moderate $\mathsf{IBO}$]{\includegraphics[scale = 1.1]{spectral_mask_ref.eps}} 
   \subfloat[high $\mathsf{IBO}$]{\includegraphics[scale = 1.1]{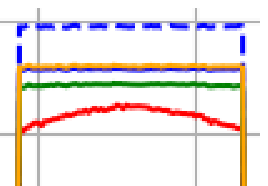}}
\end{tabular}
\caption{Spectral mask behavior trade-off between IBO and $\lambda_{2b}$.\vspace{-0.6cm}}
\label{fig:spec_behave}
\end{figure}

The proposed $\mathsf{CAE}$ decreases the out-of-band distortions at the expense of lower transmitted power efficiency.
Observing the spectral behavior as a part of the experimental analysis showed us that there is a trade-off between increasing the $\mathsf{IBO}$ and increasing the Lagrange multiplier associated with the $\mathsf{PAPR}$ loss component, $\lambda_{2b}$. As shown in Fig. \ref{fig:spec_behave}, while increasing the $\mathsf{IBO}$ mostly shifts the $\mathsf{CAE}$ curve downwards, increasing $\lambda_{2b}$ causes the curve to bend more.
\vspace{-0.3cm}
\vspace{-0.1cm}
\vspace{-0.6cm}
The transmitter's $\mathsf{OBO}$, which evaluates the power efficiency of the system, is defined as the ratio between the maximal radiated power that is the maximal power transmitted by all the $\mathsf{HPA}$s in the network, $P_T$, and the mean transmitted power at the $\mathsf{HPA}$s' input, i.e.
\begin{align}
\vspace{-0.3cm}
    \mathsf{OBO} = \frac{P_T}{\sum_{m=1}^{N_t}\mathbb{E}\left(|{{x}_{n,m}^\s{B}}|^2\right)}. 
\end{align}
\begin{table}[t]
\begin{center}
\vspace{-0.3cm}
\caption{ACPR and OBO}
\begin{tabular}{c||c|c|c|c|c||c|c|c|c|c}
     & \multicolumn{5}{c||}{\textbf{4QAM 2X2 MIMO}} & \multicolumn{5}{c}{\textbf{16QAM 4X4 MIMO}}  \\\hline
     Parameter&CAE&FC-AE&CF&SLM&No-reduction &CAE&FC-AE&CF&SLM&No-reduction  \\\hline\hline
     ACPR[dB]& -39.87 & -37.26 & -39.08 & -37.73  & -34.99 & -37.88 & -36.53 & -39.004 & -37.67  & -35.01\\\hline
     OBO[dB]& 5.92 & 6.62 & 6.74 & 6.78 & 6.86 & 6.09 & 6.77 & 6.74 & 6.78 & 6.84
\end{tabular}
 \vspace{-1.6cm}
\label{ACPR_OBO}
\end{center}
\end{table}
\begin{wrapfigure}{r}{0.5\textwidth}
\centering
\vspace{-2cm}
    \includegraphics[scale = 0.55]{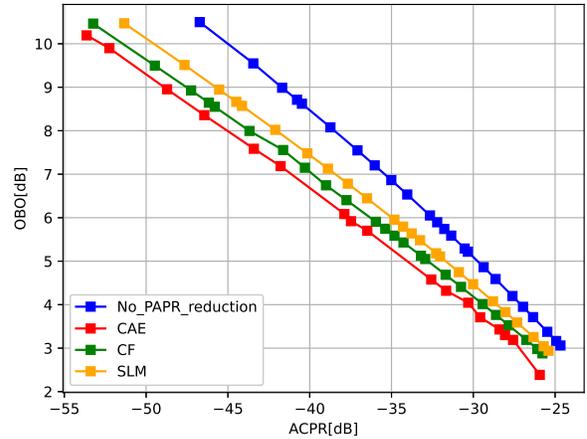}
    \caption{OBO vs. ACPR for the considered methods for 16-QAM with 4x4 MIMO setup and 3GPP multipath channel.\vspace{-0.7cm}}
    \label{fig:ACPR_vs_OBO}
\end{wrapfigure}
\vspace{-0.3cm}
The maximum radiated power is defined to be $P_T = 1$. $P_T$ is divided equally between the $\mathsf{HPA}$s. In other words, the maximal transmitted power of each $\mathsf{HPA}$ is $P_{T}/N_{t}$. As we defined all $\mathsf{HPA}$s to have the same characteristics, the saturation level of each $\mathsf{HPA}$ is $A_0=\sqrt{P_{T}/N_{T}}$.     
Table \ref{ACPR_OBO} compares the $\mathsf{ACPR}$ and the $\mathsf{OBO}$ of the proposed $\mathsf{CAE}$ to the other methods. As shown, the $\mathsf{ACPR}$ of the $\mathsf{CAE}$ is comparable with the considered methods.

In Fig. \ref{fig:ACPR_vs_OBO} we further compare the $\mathsf{OBO}$ performance for different $\mathsf{ACPR}$ values.  
It can be seen that the $\mathsf{CAE}$ system requires lower $\mathsf{OBO}$s, which is better overall power efficiency, while maintaining better $\mathsf{BER}$ compared to the other methods.
\vspace{-0.4cm}
\subsection{Autoencoder - FC vs. CNN}
\begin{figure}[t]
     \centering
     \begin{subfigure}[b]{0.49\textwidth}
         \centering
         \includegraphics[width=\textwidth]{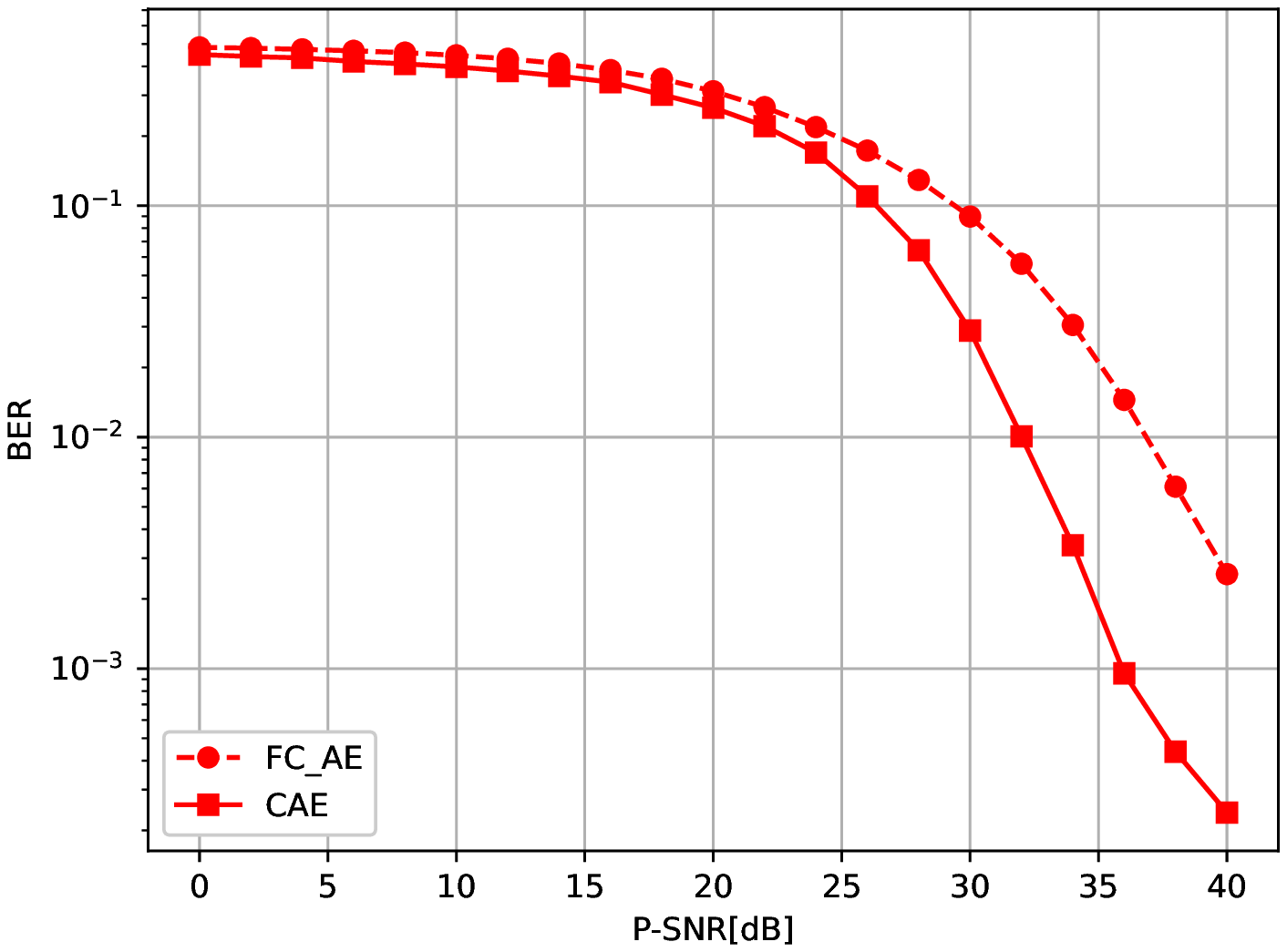}
         \caption{FC-AE and CAE}
         \label{fig:BER_comp}
     \end{subfigure}
     \hfill
     \begin{subfigure}[b]{0.49\textwidth}
         \centering
         \includegraphics[width=\textwidth]{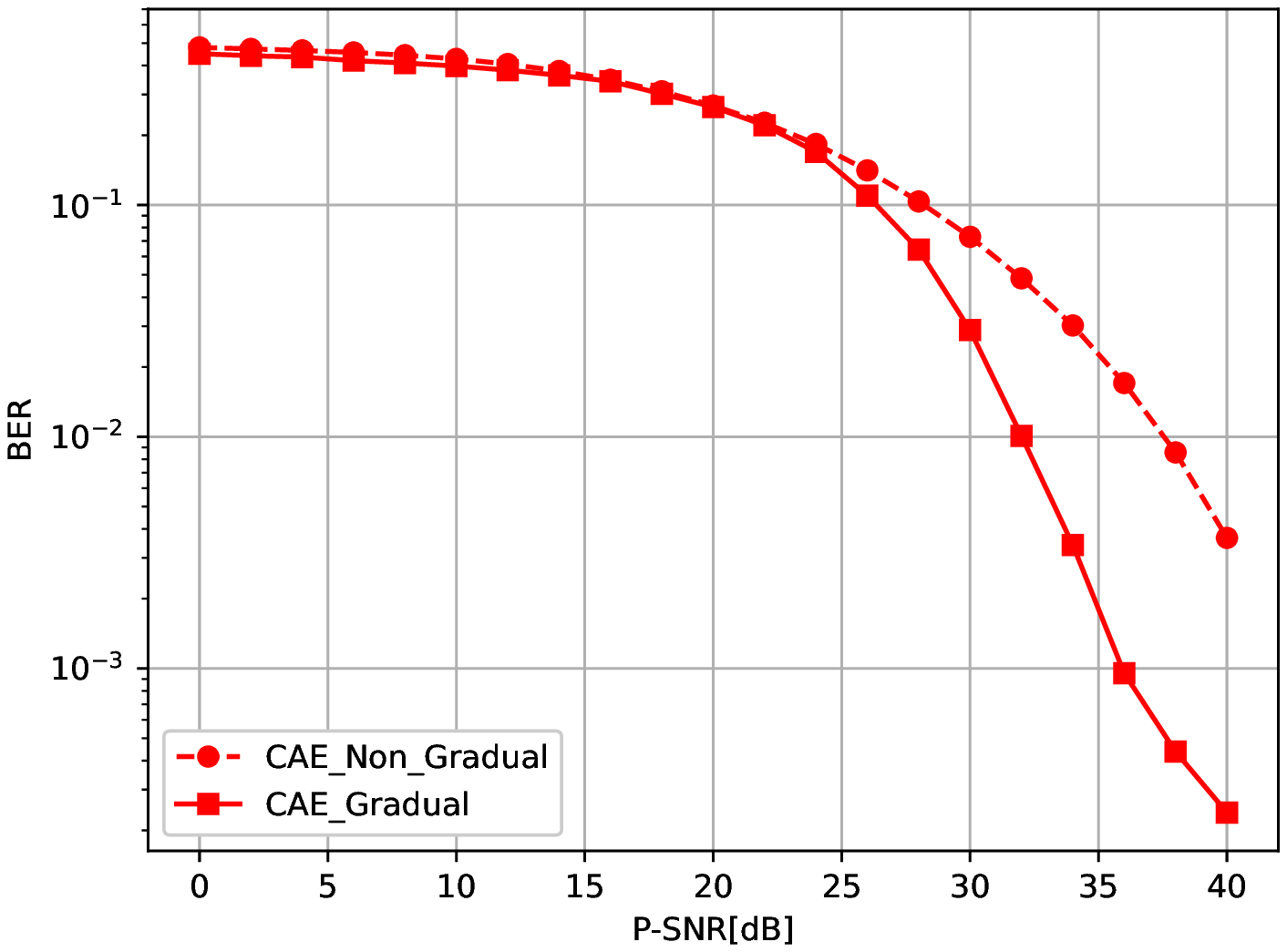}
         \caption{Fixed and gradual loss learning}
         \label{fig:BER_comp_grad}
     \end{subfigure}
        \caption{BER vs. $\s{P\_SNR}$ learning approaches comparison of 16-QAM, 4x4 MIMO with 3GPP multipath channel.\vspace{-1.1cm}}
        \label{fig:two ber graphs}
\end{figure}

\vspace{-0.1cm}
We investigated various $\mathsf{NN}$ types for the $\mathsf{AE}$, in particular, $\mathsf{FC}$ and $\mathsf{CNN}$. Figure \ref{fig:BER_comp} compares the $\mathsf{BER}$ performance of two $\mathsf{AE}$ architectures: the proposed $\mathsf{CAE}$, which contains convolutional layers, and a fully connected autoencoder (FC-AE), which contains only $\mathsf{FC}$ layers. 
It can be observed that the $\mathsf{CAE}$ network has better $\mathsf{BER}$ performance compared to the FC-AE. 
As shown in Table \ref{ACPR_OBO}, the $\mathsf{ACPR}$ of the $\mathsf{CAE}$ is better than that of the FC-AE.
Moreover, the $\mathsf{CAE}$ has lower complexity and thus faster training. 
The three convolutional layers have a total of $1953$ parameters, while for three $\mathsf{FC}$ layers of sizes $3500$, $2500$, and $3500$, as were used for the FC-AE in Fig. \ref{fig:BER_comp} and Table \ref{ACPR_OBO}, the number of parameters is around $10^7$.
\vspace{-0.8cm}
\subsection{Fixed vs. Gradual Loss Learning}
\vspace{-0.4cm}
To show the benefits of using a gradual loss learning procedure, Fig. \ref{fig:BER_comp_grad} compares its $\mathsf{BER}$ performance to that of a fixed-loss training procedure, where the loss function's weights are fixed for the entire training. It can be observed that the gradual loss learning procedure significantly improves the $\mathsf{BER}$. In addition, improving the $\mathsf{BER}$ while keeping the $\mathsf{PAPR}$ and spectral performance at the desired levels is easier to control when applying the gradual loss learning method than manipulating loss function weights in fixed-loss training. Also, spectral performance and $\mathsf{PAPR}$ reduction were harder to control and provide similar performance.
\vspace{-0.4cm}
\section{Conclusions and Future Work}
\vspace{-0.2cm}
In this study, we have presented a $\mathsf{CAE}$ model for $\mathsf{PAPR}$ reduction and waveform design in a $\mathsf{MIMO}$-$\mathsf{OFDM}$ communication system. We have applied a gradual loss learning method to optimize the performance in terms of three objectives: low $\mathsf{BER}$, low $\mathsf{PAPR}$, and adherence to $\mathsf{ACPR}$ spectral requirements, on top of the $\mathsf{AL}$ multipliers optimization technique. The presented $\mathsf{CAE}$ structure trainable parts included a neural $\mathsf{PAPR}$ reduction block, followed by a $\mathsf{BPF}$ filter to optimize the spectral behavior at the transmitter, and a neural iterative $\mathsf{MIMO}$ detection block at the receiver, both were simultaneously optimized as a part of the end-to-end network design. The proposed $\mathsf{CAE}$ was shown to outperform the $\mathsf{CF}$ and the $\mathsf{SLM}$ algorithms at the examined cases. Future work can extend the $\mathsf{MIMO}$ scenario to higher modulation schemes and larger $\mathsf{MIMO}$ setups, aiming to achieve a functional utility for future wireless communication networks.
\vspace{-1cm}
\bibliography{ref}
\bibliographystyle{IEEEtran}

\end{document}